\documentclass{aa}  
\usepackage[colorlinks=true, linkcolor=blue, citecolor=blue, urlcolor=blue]{hyperref}

\usepackage{graphicx}
\usepackage{txfonts}
   \title{Exploring the Sagittarius stream with RR Lyrae Stars from Gaia Data Release 3}
\begin{document}

   \subtitle{}

   \author{T. Muraveva
          \inst{1}
          \and
          M. Bellazzini\inst{1}
          \and 
          A. Garofalo\inst{1}
          \and
          G. Clementini\inst{1}
          \and 
          L. Monti\inst{1}
          \and
          M. L. Valentini\inst{2}
          }

   \institute{INAF - Osservatorio di Astrofisica e Scienza dello Spazio di Bologna, Via Piero Gobetti 93/3, Bologna 40129, Italy\\
              \email{tatiana.muraveva@inaf.it}
   \and Dipartimento di Fisica e Astronomia, Alma Mater Studiorum - University of Bologna, Via Piero Gobetti 93/2,  Bologna 40129, Italy
              }

   \date{Received ; accepted }

% \abstract{}{}{}{}{} 
% 5 {} token are mandatory

  \abstract
   {The Sagittarius (Sgr) dwarf spheroidal galaxy is one of the most prominent satellites of the Milky Way (MW). It is currently undergoing tidal disruption, forming an extensive stellar stream that provides key insights into the assembly history of the MW halo.} 
    % The Sagittarius (Sgr) dwarf spheroidal galaxy is one of the most prominent satellites of the Milky Way (MW). It is currently undergoing tidal disruption, forming an extensive stellar stream that provides key insights into the assembly history of the MW halo.
   {Our goal is to investigate the structure and metallicity distribution of the Sgr stream using RR Lyrae stars (RRLs).}
 {We analyzed RRLs provided in {\it Gaia} Data Release 3 (DR3), for which new estimates of photometric metallicities are available in the literature, and accurate distances were calculated using the reddening-free period-Wesenheit-metallicity ($PWZ$) relation in the {\it Gaia} $G$, $G_{BP}$, and $G_{RP}$ bands.}
{We determine the mean metallicity of RRLs in the Sgr stream to be ${\rm [Fe/H]}=-1.62 \pm 0.01$~dex.  We measure a metallicity gradient as a function of stripping time from the Sgr progenitor of $0.05 \pm 0.02$ dex/Gyr, indicating that the metal-poor RRLs were stripped earlier during the accretion process. The far arm is found to be the most metal-poor structure of the Sgr stream, with a mean metallicity of ${\rm [Fe/H]}=-1.98 \pm 0.37$~dex, significantly lower than that of the leading ($-1.69\pm0.31$~dex) and trailing ($-1.64 \pm 0.28$~dex) arms. Our findings show that the RRLs in the far arm of the Sgr stream exhibit a bimodal metallicity distribution with peaks at [Fe/H]=$-2.4$~dex and $-1.7$~dex. The main body of the stream is the most metal-rich structure, with a mean metallicity of ${\rm [Fe/H]}=-1.58 \pm 0.31$~dex and a radial gradient of $-0.008 \pm 0.005$~dex/kpc. We find almost negligible metallicity gradients of $(-0.2 \pm 0.3)\times 10^{-3}$ dex/deg in the trailing arm and $(-1.0 \pm 0.5)\times 10^{-3}$ dex/deg in the leading arm, in agreement with previous studies. Finally, we investigate the bifurcation of the Sgr stream and conclude that the metallicity difference between the faint and bright branches is not confirmed based  on the RRLs in our sample.}
{}

\keywords{Stars: variables: RR Lyrae - stars: abundances - galaxy: halo - surveys - techniques: photometric}

\maketitle
%
%-------------------------------------------------------------------

\section{Introduction}\label{sec:intr}

The Sagittarius (Sgr) dwarf spheroidal galaxy  (dSph) is one of the most significant and well-studied examples of ongoing galactic accretion within the Milky Way (MW), offering key insights into the hierarchical assembly of our Galaxy. First discovered by \citet{Ibata1994}, Sgr dSph is in the final stages of being tidally disrupted by the MW’s gravitational field. This disruption has given rise to an extensive stellar stream that envelops the MW providing a direct observational record of the merger process \citep{Majewski2003, Belokurov2006, belok14,Ibata2020,Antoja2020}. 

Like any tidal tail, the Sgr stream consists of a leading arm that precedes and a trailing arm that lags behind the main body of the galaxy along its polar orbit.
%As with any tidal tail, the Sgr stream consists of a leading and a trailing arm, which precede and lag behind the main body of the galaxy, respectively, along its polar orbit.
A metallicity gradient along the stream has been reported over the past two decades \citep{bellaz06,chou07,monaco07}, with the prevalence of metal-poor stars increasing with distance from the main body. This trend is generally interpreted as the result of tidal disruption acting on a pre-existing gradient in the Sgr progenitor at the time of infall \citep[see, e.g.,][and references therein]{deboer14, gibbons17, Yang2019, Hayes2020, Ramos2022, minelli23}. Numerous studies have attempted to model the Sgr stream, aiming to reconstruct the system’s dynamical evolution and to use the stream to constrain the properties of the MW dark matter halo (e.g., \citealt{Ibata1998}; \citealt{Helmi2001}; \citealt{Law2010}; \citealt{Dierickx2017}; \citealt{Thomas2017}; \citealt{Fardal2019}; \citealt{Oria2022}), including the influence of the Large Magellanic Cloud (LMC) on the merger \citep{Vera-Ciro2013, Vasiliev2021}.

Among the stellar populations in the Sgr dSph and its stream, RR Lyrae stars (RRLs) are of particular interest. Although recent studies suggest that relatively young and metal-rich RRLs can form through the evolution of close binary systems \citep{Bobrick2024}, the majority of RRLs are old (age $\geq 10$~Gyr), metal-poor, pulsating variables that populate the MW halo and old stellar systems such as globular clusters (GCs) and dSphs. This makes them highly effective tracers of structures associated with past accretion events, especially because: (1) they are excellent distance indicators \citep{catelan04}; and (2) their metallicity can be estimated from photometric parameters, such as the pulsation period and Fourier decomposition parameters of their light curves, without requiring spectroscopic data (e.g., \citealt{Jurcsik1996}; \citealt{Morgan2007}). Individual distances to RRLs can be determined using, for instance, the absolute magnitude–metallicity relation in the visual band ($M_V$–[Fe/H]; e.g., \citealt{Clementini2003}; \citealt{Bono2003}), or in the {\it Gaia} $G$ band ($M_G$–[Fe/H]; e.g., \citealt{Muraveva2018}, \citealt{Li2023}),  near-/mid-infrared period–luminosity–metallicity ($PLZ$) relations (e.g., \citealt{Longmore1986}; \citealt{Sollima2008}; \citealt{Madore2013}; \citealt{Muraveva2015, Muraveva2018}; \citealt{Neeley2017, Neeley2019}; \citealt{Prudil2024}); and the period–Wesenheit–metallicity ($PWZ$) relation in the $G$, $G_{BP}$, and $G_{RP}$ bands (e.g., \citealt{Garofalo2022}; \citealt{Li2023}; \citealt{Prudil2024}).

%Thus, RRLs can be used to measure the distances to different substructures of the Sgr stream and reconstruct its three-dimensional geometry. Moreover, the metallicity of RRLs can be estimated based solely on photometric parameters (e.g., \citealt{Jurcsik1996}; \citealt{Morgan2007}), making them useful for studying the metallicity distribution in the Sgr stream. Additionally, RRLs, being typically old and metal-poor, are expected to dominate the outskirts of dSphs and to be the first stars stripped from the progenitor, making them valuable for studying the oldest components of the Sgr stream. Finally, as an old population, RRLs reveal the metallicity gradient in the Sgr progenitor as it was before being accreted by the MW.

Indeed, RRLs have been widely used to study the Sgr stream. \citet{Hernitschek2017} analyzed RRLs observed by the Pan-STARRS1 survey (PS1; \citealt{Kaiser2010}) to trace the leading and trailing arms of the stream and to characterize their distances and line-of-sight depths. \citet{Sesar2017} used the same RRLs dataset to trace the trailing arm out to its apocenter, at a distance of about 90~kpc, where it bifurcates into two components: one bending toward the Galactic center and an extra arm extending to 120~kpc, referred to as the “spur.” The latter feature was recently studied in more detail by \citet{Bayer2025} using blue horizontal branch (BHB) stars.

%Additionally, using the sample of RRLs, \citet{Sesar2017} clearly show that there is an extension of the trailing arm up to distance of 90 kpc, initially discovered by \citet{Newberg2003}. We refer to this structure as the "far arm" in our study.

%The study of RRLs in the Sgr system has accelerated in recent years with the arrival of data from the {\it Gaia} mission \citep{Prusti2016}. 

With the advent of the {\it Gaia} Data Release 2 (DR2; \citealt{Brown2018}), the Early Data Release 3 (EDR3, \citealt{Brown2021}) and the Data Release 3 (DR3; \citealt{Vallenari2023}), which include high-precision measurements of positions, parallaxes, and proper motions for almost two billion stars, along with the identification and characterization of thousands of RRLs (\citealt{Clementini2019, Clementini2023}; \citealt{Rimoldini2019}), it has become possible to study the distribution, kinematics, and metallicity of RRLs in the Sgr stream with unprecedented detail. \citet{Ramos2020} used RRLs from {\it Gaia} DR2 (\citealt{Brown2018}, \citealt{Clementini2019}) to study the 5-D distribution and metallicity gradient of the Sgr stream. Similarly, \citet{Ibata2020} used RRLs from the {\it Gaia} DR2 catalogue and their photometric metallicities to provide distance measurements along the stream using the $M_G-{\rm [Fe/H]}$ relation from \citet{Muraveva2018}. However, photometric metallicities provided in the {\it Gaia} catalogue \citep{Clementini2019, Clementini2023}, calibrated using the relations from \citet{Nemec2013}, have been found to be biased toward higher metallicities (see discussion in \citealt{Muraveva2025}), which may lead to underestimated distances and possibly help explain the discrepancy between the distances from the \citet{Law2010} model and those derived from RRLs by \citet{Ibata2020}.

\citet{Antoja2020} used the {\it Gaia} DR2 astrometry to detect the Sgr stream from proper motions alone and, for the first time, determined the proper motion along the entire path of the stream. \citet{Ramos2022} constructed the largest sample of more than 700,000 candidate members of the Sgr stream among which 8060 RRLs, using {\it Gaia} EDR3 \citep{Brown2021}. \citet{Wang2022} identified 145 RRLs as belonging to the Sgr stream based on six-dimensional position-velocity information from the Sloan Digital Sky Survey (SDSS, \citealt{York2000}) Sloan Extension for Galactic Understanding and Exploration (SEGUE; \citealt{Yanny2009}) and the Large Sky Area Multi-Object Fiber Spectroscopic Telescope (LAMOST; \citealt{Cui2012}), as well as the Gaia EDR3. Recently, \citet{Sun2025} used a sample of RRLs from the {\it Gaia} DR3 (\citealt{Vallenari2023}, \citealt{Clementini2023}) catalogue, for which photometric metallicities and distances were measured by \citet{Li2023} to study different Galactic substructures, including the Sgr stream. 

In this paper, we revise the geometric properties of the Sgr stream and the metallicity of its oldest stellar component using RRLs from the {\it Gaia} DR3 catalogue as tracers, with three key differences compared to previous studies: (1) we use the new photometric metallicities from \citet{Muraveva2025}, derived by applying machine learning techniques to the {\it Gaia} light curves of $\sim135,000$ RRLs; (2) we obtain new distance estimates for these RRLs by incorporating the new metallicities into the reddening-free $PWZ$ relation from \citet{Garofalo2022}, after verifying that there is no residual trend between metallicity and distance; and (3) we use three different samples of candidate stream members with varying degrees of purity and completeness, both for cross-validation and to extend the analysis of the metallicity distribution to more distant regions of the stream that have not been previously studied from this perspective.

The paper is structured as follows: Section~\ref{sec:data} describes the dataset used in our study and the selection procedures adopted to construct our samples of Sgr RRLs. In Section~\ref{sec:met}, we analyze the metallicity distribution in the Sgr trailing, leading, and faint arms, as well as the main body. Section~\ref{sec:summ} summarizes our main results.

%--------------------------------------------------------------------
\section{Data}\label{sec:data}

\subsection{{\it Gaia} DR3 sample of RRLs}\label{sec:gaia}

%\begin{figure*}
%\includegraphics[width=19cm]{figures/XZ2.png}
%\caption{Density distribution on the Cartesian $X-Z$  plane of the metal-poor ([Fe/H]<-2 dex, {\it left panel}), intermediate metallicity (-2<[Fe/H]<-1 dex, {\it medium panel}) and  metal-rich ([Fe/H]>-1 dex, {\it bottom panel}) RRLs.}\label{fig:xyz_met}\label{fig:xz}
%\end{figure*}

\begin{figure*}
\includegraphics[width=6cm]{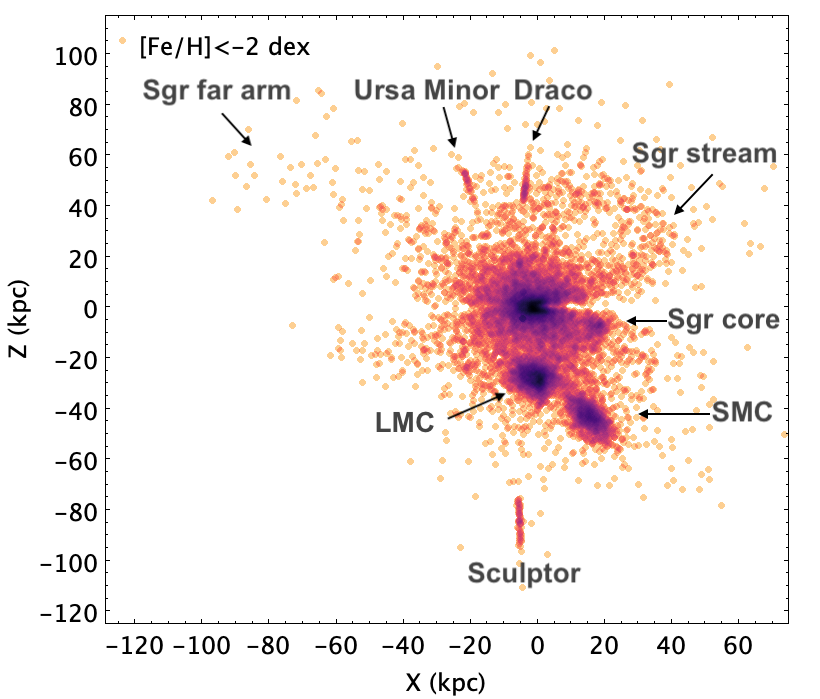}
\includegraphics[width=6cm]{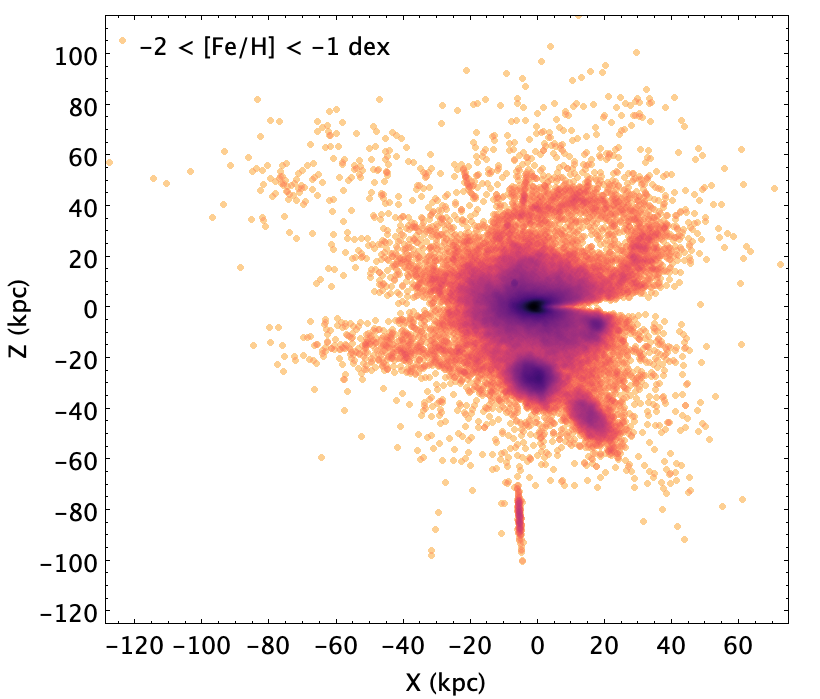}
\includegraphics[width=6cm]{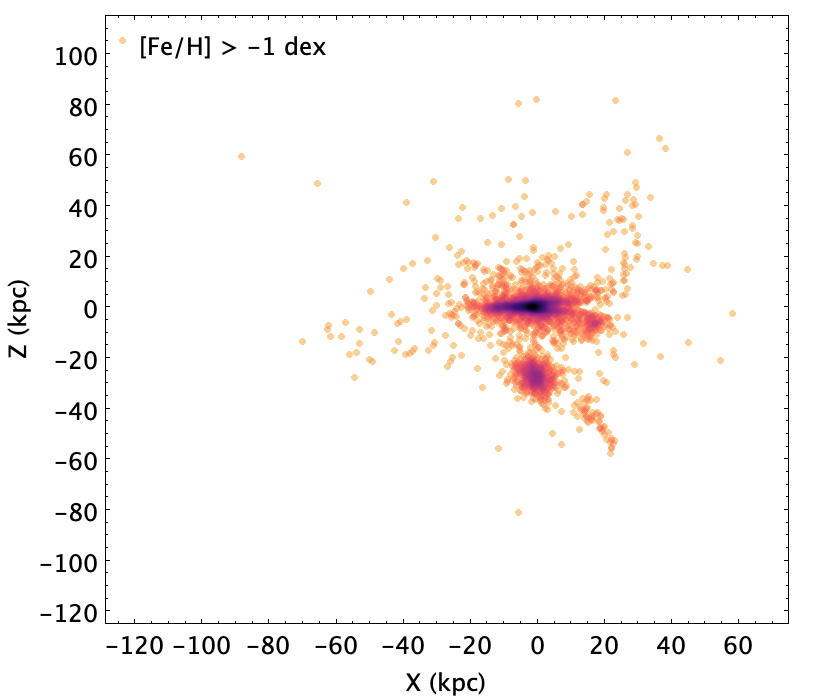}
\caption{Distribution on the Cartesian $X-Z$  plane of metal-poor ([Fe/H]<$-2$ dex, {\it left panel}), intermediate metallicity ($-2$<[Fe/H]<$-1$ dex, {\it medium panel}) and  metal-rich ([Fe/H]>$-1$ dex, {\it right panel}) RRLs.}\label{fig:xyz_met}\label{fig:xz}
\end{figure*}

The {\it Gaia} DR3 includes a clean catalogue of 270,891 RRLs analyzed by the Specific Object Study (SOS) pipeline for Cepheids and RRLs (SOS Cep\&RRL; \citealt{Clementini2023}). The catalogue provides, among other parameters, pulsation periods, peak-to-peak amplitudes of the $G$, $G_{BP}$, and $G_{RP}$ light curves, mean magnitudes, and Fourier decomposition parameters of the $G$-band light curves. In \citet{Muraveva2025}, we cleaned the sample of RRLs from the {\it Gaia} DR3 catalogue by comparing it with the OGLE IV results \citep{OGLE1, OGLE2} and analyzing the distribution of RRLs on the Bailey (amplitude in the $G$ band versus pulsation period) diagram, resulting in a final sample of 258,696 RRLs. We then presented new relations between the metallicities of RRLs, their pulsation periods, and the Fourier decomposition parameters published in {\it Gaia} DR3. These relations were calibrated using accurate spectroscopic metallicities available in the literature \citep{Crestani2021, Liu2020}. A feature selection algorithm was applied to identify the most relevant parameters for determining metallicity. To fit the relations, we used a Bayesian approach, accounting for uncertainties in the parameters and the intrinsic scatter of the relations. As a result, photometric metallicity estimates were derived for 134,769 RRLs from the cleaned {\it Gaia} DR3 sample.

Using the photometric metallicities from \citet{Muraveva2025}, we calculated distances from the Sun ($D_\odot$) for 133,767 RRLs in our sample by applying the reddening-free $PWZ$ relation in the {\it Gaia} $G$, $G_{BP}$, and $G_{RP}$ bands provided by \citet{Garofalo2022} (their eq. 21). This relation was calibrated using a hierarchical Bayesian approach and precise parallaxes of field RRLs from the {\it Gaia} EDR3 catalogue. The apparent Wesenheit magnitudes were calculated following the formulation by \citet{Ripepi2019}:
\begin{equation}\label{eq:lambda}
W(G, G_{BP}, G_{RP}) = G - \lambda(G_{BP} - G_{RP}),
\end{equation}
where $\lambda = A(G)/E(G_{BP} - G_{RP})$. Following \citet{Garofalo2022}, we adopted the value $\lambda = 1.922 \pm 0.045$. Mean $G$ magnitudes, computed as intensity averages over the full pulsation cycle, were obtained from the {\it Gaia} DR3 catalogue of RRLs (\texttt{vari\_rrlyrae} table, \citealt{Clementini2023}), while $G_{BP}$ and $G_{RP}$ mean magnitudes were taken from the general {\it Gaia} DR3 catalogue (\texttt{gaia\_source}, \citealt{Vallenari2023}).

The $PWZ$ relation from \citet{Garofalo2022} is on the \citet{ZW1984} metallicity scale, while the metallicities provided by \citet{Muraveva2025} are on the scale adopted by \citet{Crestani2021}. To ensure consistency, we first transformed the photometric metallicities from \citet{Muraveva2025} to the \citet{Carretta2009} scale by subtracting 0.08 dex, as recommended by \citet{Crestani2021} and \citet{Mullen2021}. We then converted these metallicities to the \citet{ZW1984} scale using the equation from \citet{Carretta2009}:
\begin{equation}\label{eq:zw-carretta}
{\rm [Fe/H]_{ZW84}} = ({\rm [Fe/H]_{C09}} - 0.160)/1.105.
\end{equation}

Uncertainties in our distance estimates arise from errors in individual metallicity values, apparent $G$, $G_{BP}$, and $G_{RP}$ mean magnitudes, the $\lambda$ value, and the coefficients of the $PWZ$ relation, as well as its intrinsic scatter ($\sigma = 0.09$~mag). We estimated uncertainties in the distances using a Monte Carlo simulation approach. For each star, 1000 iterations were performed, with random values sampled from the error distributions of metallicities, apparent mean magnitudes, $\lambda$ value, and the coefficients of the $PWZ$ relation, while also simulating its intrinsic dispersion. Collecting the mean distance values and their standard deviation obtained from the Monte Carlo simulation allowed us to estimate uncertainties in distances. RRLs with relative distance errors exceeding 15\% or metallicity errors greater than 1 dex were removed, resulting in a clean sample of 131,274 RRLs. We then calculated the Galactocentric Cartesian coordinates ($X$, $Y$, $Z$) for the RRLs in our clean sample using their positions and estimated distances. The position of the Sun was assumed to be on the X-axis of a right-handed coordinate system. The X-axis points from the position of the Sun to the Galactic centre, while the Y-axis points towards Galactic longitude $l=90^{\circ}$. The Z-axis points towards the North Galactic Pole ($b = 90^{\circ}$). The Sun was assumed to be at a distance of 8.122~kpc from the Galactic centre \citep{Gravity2018}.

Figure \ref{fig:xyz_met} shows the distribution of metal-poor ([Fe/H] < $-2$ dex), metal-rich ([Fe/H] > $-1$ dex), and RRLs with metallicities ranging from $-2$ to $-1$ dex on the Cartesian $X-Z$ plane.  Well-known structures, such as the LMC, the Small Magellanic Cloud (SMC), the Sgr stream, and the main body of the Sgr dSph are clearly visible, demonstrating the potential of using RRLs to study the structures and metallicity distributions in the MW and beyond. The dSphs, such as Draco, Ursa Minor, and Sculptor, are well-resolved in the metal-poor and intermediate-metallicity regimes. As expected, these dSphs do not contain metal-rich RRLs, which are mainly distributed in the Disk of the MW. Notably, the Sgr stream is fully traced by metal-poor and intermediate-metallicity RRLs, while the metal-rich RRLs are confined to the inner part of the leading arm, near the main body and within the main body itself. This provides insights into the evolution of the Sgr stream. 

\begin{figure*}
\includegraphics[width=19cm]{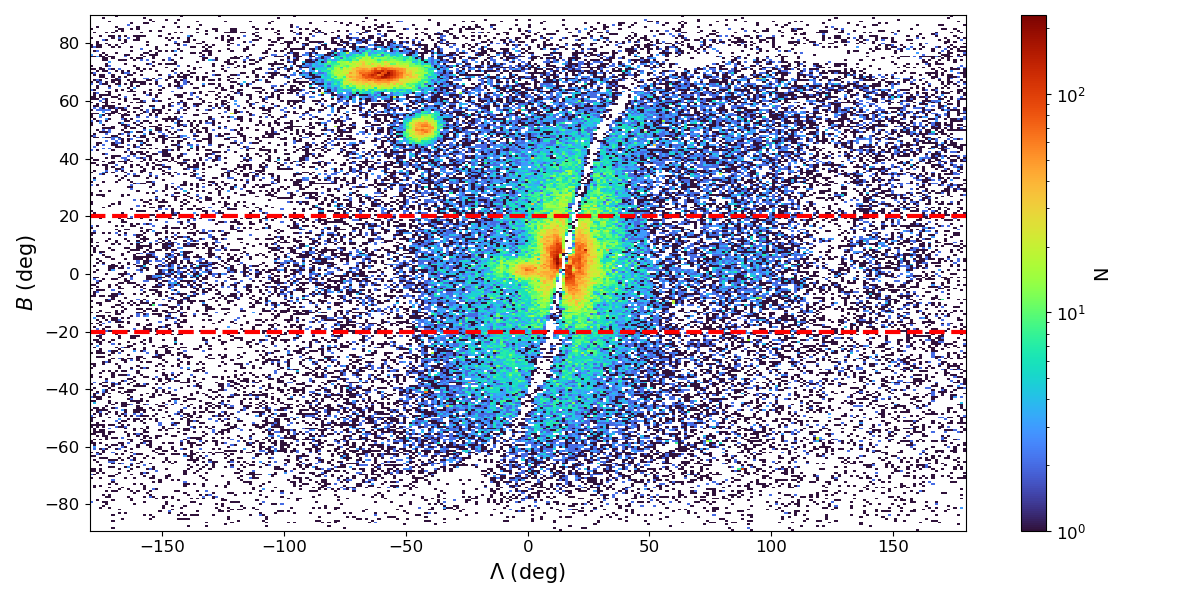}
\caption{Density map of the 131,274 RRLs from the clean sample in the Sgr coordinate system ($B$ versus $\Lambda$), as defined by \citet{Vasiliev2021}. The red dashed lines correspond to $B=20^\circ$ and $B=-20^\circ$.}\label{fig:lambda_beta}
\end{figure*}

We use the \texttt{gala} Python package \citep{gala} to transform the coordinates of RRLs in our clean sample to the coordinate system $\Lambda$ and $B$ defined by the orbit of the Sgr dSph as presented by \citet{Vasiliev2021}. 
The $\Lambda$ values increase towards the leading arm and are equal to zero at the center of the Sgr remnant. 
The orbital pole of the Sgr plane has $B = 90^\circ$ ($l = 273^\circ.75$, $b = -13^\circ.46$ in Galactic coordinates). 
%\citet{Vasiliev2021} used a standard right-handed Galactocentric coordinate system as defined in the Python library \texttt{ASTROPY} \citep{Astropy}, in which the Sun is located at $x_{\odot} = \{-8.1, 0, 0.02\}$ kpc and moves with velocity $v_{\odot} = \{12.9, 245.6, 7.8\}$ km/s. The Galactocentric position and velocity of the Sgr remnant are 
%$x_{\rm Sgr} = \{17.9, 2.6, -6.6\}$ kpc and $v_{\rm Sgr} = \{239.5, -29.6, 213.5\}$ km/s.
Fig.~\ref{fig:lambda_beta} shows the distribution of RRLs from the clean sample on $B$ versus $\Lambda$ plane, color-coded by density. The Sgr main body and stream spanning the entire sky are clearly seen in the range of $B = [-20^\circ, 20^\circ]$. In the next section we study in more detail the RRLs belonging to the Sgr system.

\subsection{Subsets selection}\label{sec:subsets}

\begin{table*}
\caption{Subsets of RRLs in the Sgr stream used in this study}\label{tab:subsets}
\centering
\begin{tabular}{lccc}
\hline \hline
Name & N  & [Fe/H] range & Mean [Fe/H] \\
 & (RRLs) & (dex) & (dex) \\
\hline
RRLS-SGR-SELECTION & 5296 & [-3.00, 0.62] & $-1.64 \pm 0.31$ \\
RRLS-SGR-MODEL & 2377 & [-2.94, 0.62] & $-1.61 \pm 0.30$ \\
RRLS-SGR-R22 & 3865 & [-2.98, 0.62] & $-1.62 \pm 0.31$ \\
\hline
\end{tabular}
\end{table*}

\begin{figure*}
\includegraphics[width=17cm]{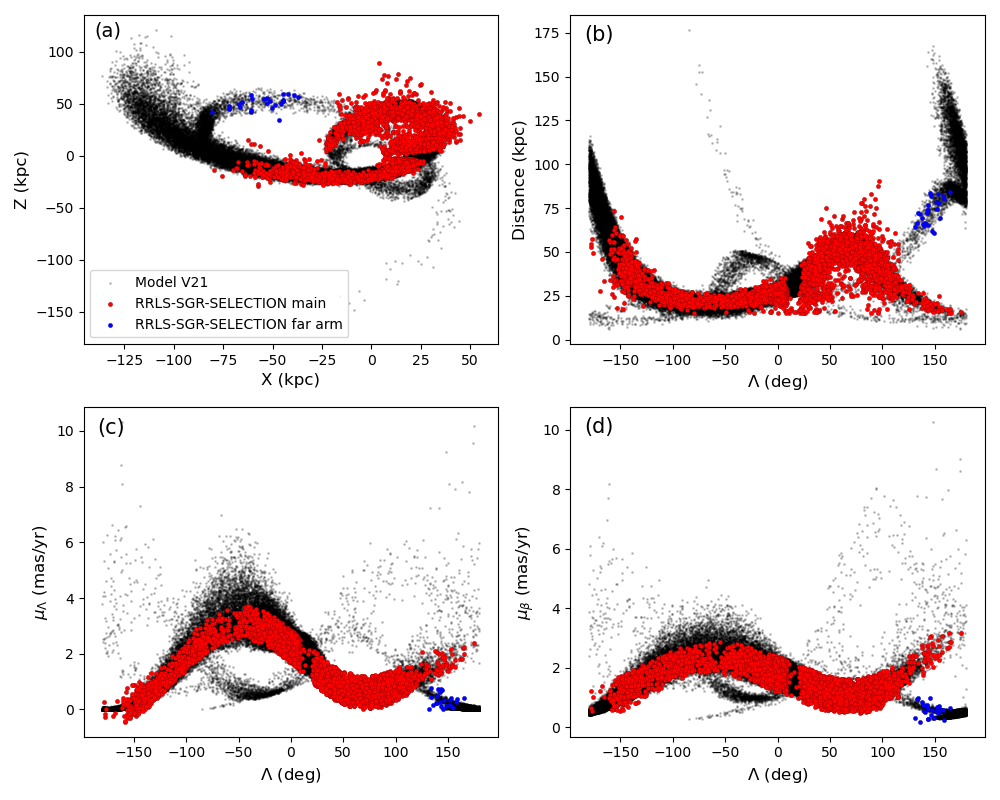}
\caption{The distributions of RRLs from the RRLS-SGR-SELECTION sample, obtained through the main selection procedure (red points) and the selection of far-arm stars as described by Eq.~\ref{eq:far} (blue points), along with simulated stars from the \citet{Vasiliev2021} model (black dots), are shown in the following planes: (a) Cartesian $Z$ versus $X$, (b) distance versus the Sgr $\Lambda$ coordinate, (c) proper motion $\mu_\Lambda$ versus $\Lambda$, and (d) proper motion $\mu_B$ versus $\Lambda$ coordinate. The Sgr coordinates and proper motions are in the system introduced by \citet{Vasiliev2021}.}\label{fig:4panel_bellazzini}
\end{figure*}

\begin{figure*}
\includegraphics[width=17cm]{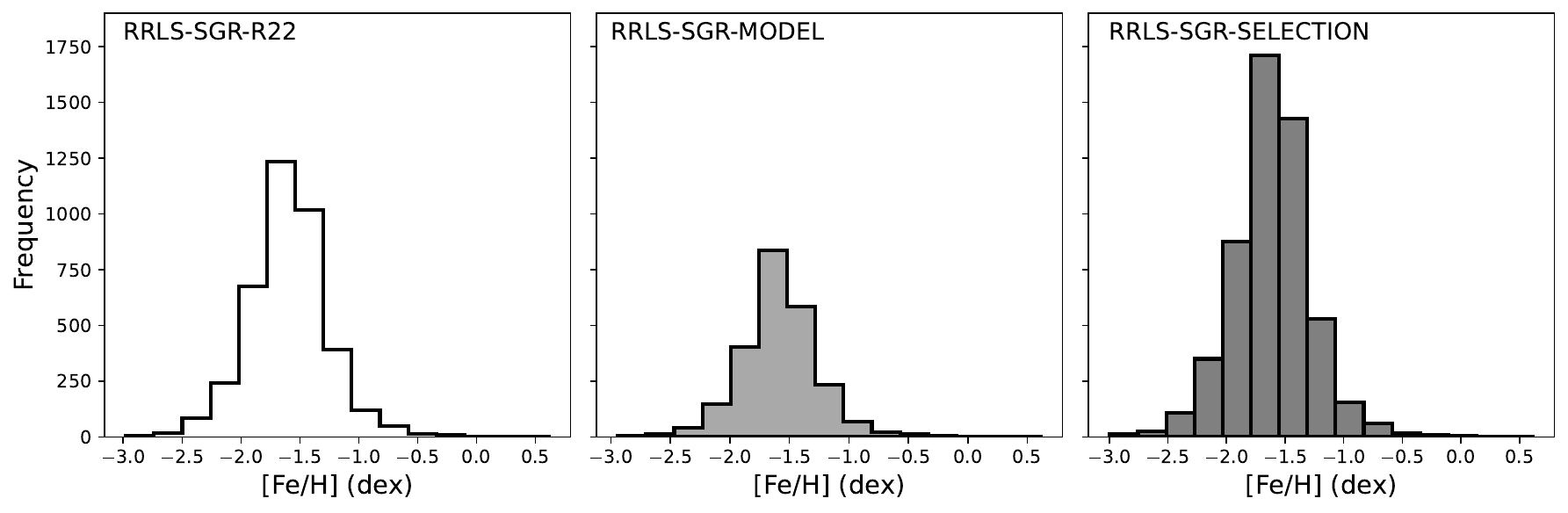}
\caption{Metallicty distributions of RRLs in the RRLS-SGR-R22, RRLS-SGR-MODEL and RRLS-SGR-SELECTION samples. \label{fig:met_hist}} 
\end{figure*}

\begin{figure*}
\includegraphics[width=17cm, trim= 80 40 200 20]{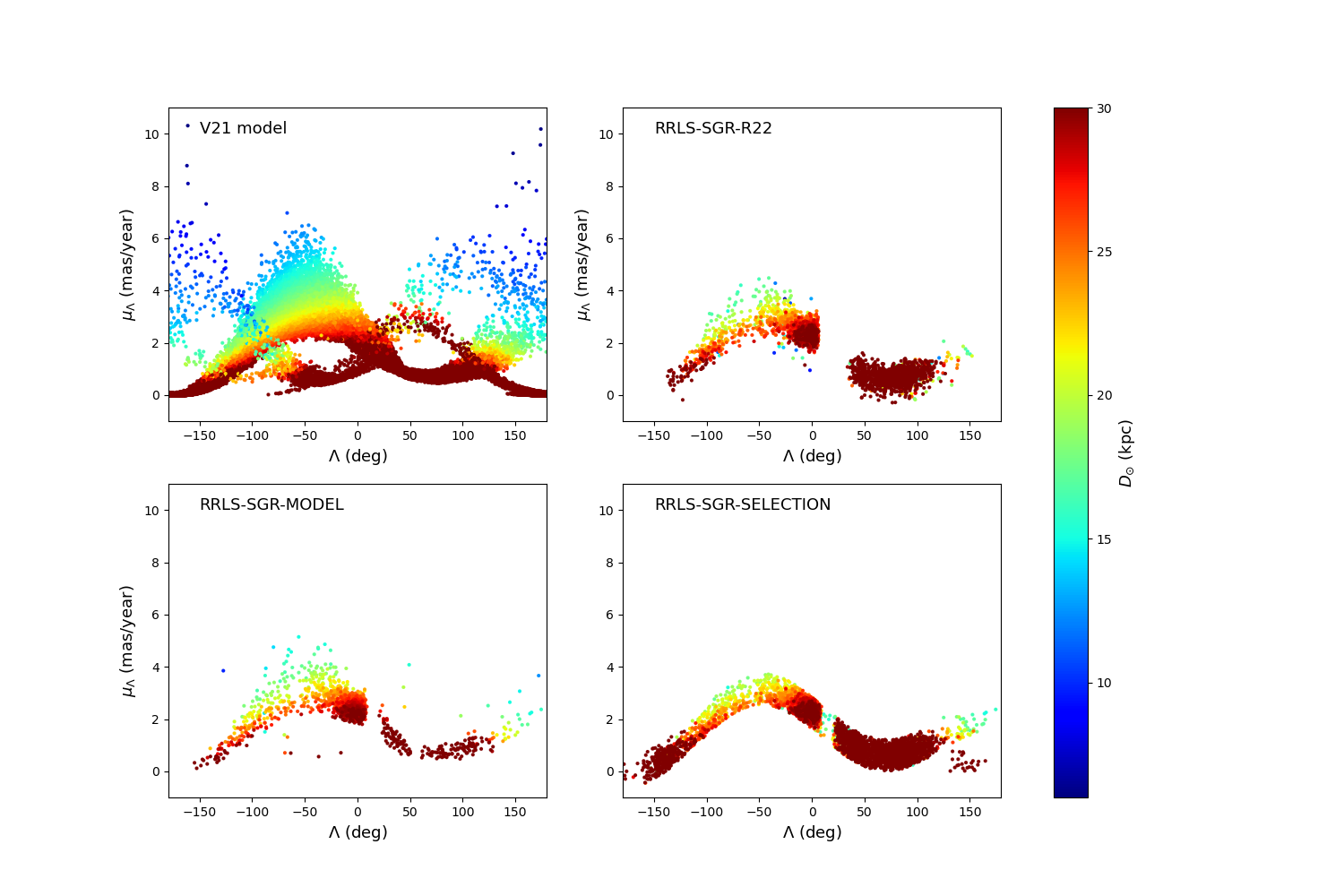}
\caption{Distributions of stars simulated by the \citet{Vasiliev2021} model and RRLs from the RRLS-SGR-MODEL, RRLS-SGR-R22, and RRLS-SGR-SELECTION samples on the $\mu_\Lambda$ versus $\Lambda$ plane, color-coded by distance. The Sgr coordinates and proper motions are given in the system introduced by \citet{Vasiliev2021}.}\label{fig:4panel_distance}
\end{figure*}

\begin{figure}
\includegraphics[width=\linewidth, trim=0 0 40 0]{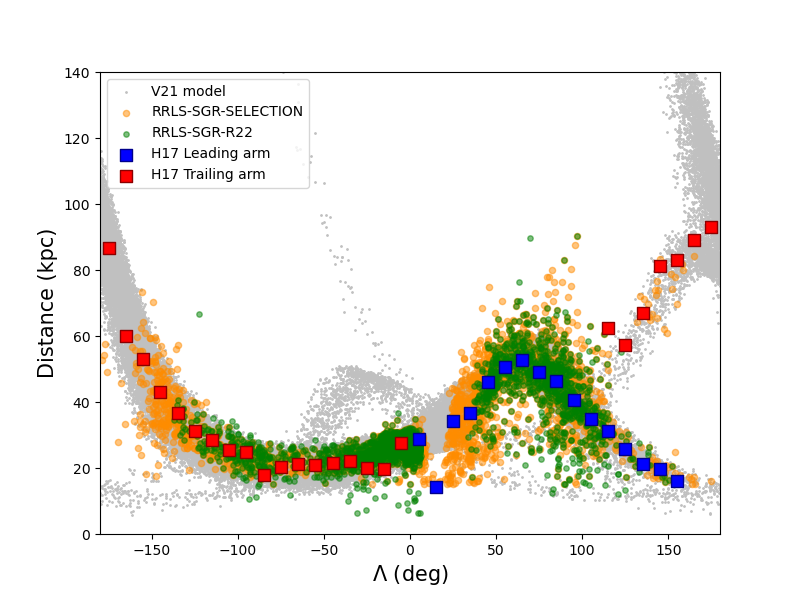}
\caption{Distributions of stars simulated by the \citet{Vasiliev2021} model (grey points), RRLs from the RRLS-SGR-R22 (green points) and RRLS-SGR-SELECTION (orange points) samples, in the distance versus $\Lambda$ plane. Blue and red squares represent the mean distances as a function of $\Lambda$ in the leading and trailing arms, respectively, as modeled by \citet{Hernitschek2017}. The Sgr $\Lambda$ coordinate is given in the system introduced by \citet{Vasiliev2021}.}\label{fig:H17_distance}
\end{figure}

We selected RRLs belonging to the Sgr stream using three different methods, obtaining three distinct subsets:

\begin{enumerate}

\item \citet{Ramos2022} used the {\it Gaia} EDR3 catalogue (\citealt{Brown2021}) to identify more than 700,000 candidate members of the Sgr stream based on strict kinematic criteria. We cross-matched our clean sample of RRLs, as defined in Section~\ref{sec:gaia}, with the sample from \citet{Ramos2022} and identified 3865 RRLs in common (Table~\ref{tab:rrls_sgr_r22}). In the following discussion, we refer to this sample as RRLS-SGR-R22. Fig.~\ref{fig:4panel_ramos} shows the distribution of RRLs in the RRLS-SGR-R22 sample.

\item \citet{Vasiliev2021} developed an $N$-body model of the disrupting Sgr galaxy within the combined gravitational potential of the MW (with a triaxial halo) and the LMC (with a mass of $1.5 \times 10^{11} M_{\odot}$). This model provides a reasonable fit to many observed properties of the Sgr system. We decided to use it as a reference to select a sample of RRLs with a 5-D phase-space distribution as similar as possible to the model's predictions. To this end, we used the publicly available snapshot of the \citet{Vasiliev2021} model at the present day, which includes 200,000 particles of the simulated Sgr system, providing their Cartesian coordinates, proper motions, distances, and stripping times, defined as the most recent time a particle left a sphere of radius 5 kpc around the Sgr progenitor.
We then performed a 5-D Cartesian cross-match between the RRLs in our clean sample and the simulated particles using the Cartesian coordinates ($X$, $Y$, $Z$) and proper motions (\texttt{pmra}, \texttt{pmdec}). The tolerance intervals for the cross-match were set as follows: $\Delta X = 2$~kpc, $\Delta Y = 2$~kpc, $\Delta Z = 2$~kpc, $\Delta \texttt{pmra} = 0.15~\mathrm{mas/yr}$, and $\Delta \texttt{pmdec} = 0.15~\mathrm{mas/yr}$, consistent with the mean errors of $2$~kpc for distances, $0.15~\mathrm{mas/yr}$ for \texttt{pmra}, and $0.13~\mathrm{mas/yr}$ for \texttt{pmdec} of the RRLs in the clean sample.

Based on this positional and kinematic match, we assume that RRLs which have a simulated counterpart likely originated from the Sgr progenitor. We also consider the stripping time of the associated model particle as a reasonable estimate of the stripping time of the matched RRL. This transfer of a label from model particles to real RRLs is not meant to be interpreted on a star-by-star basis. Rather, it is justified in a statistical sense, as particles in the model snapshot that share position and proper motion with the matched RRLs tend to have been stripped around the same epoch. Thus, this approach provides a way to rank different portions of the stream according to the timescale of tidal stripping during the disruption of the Sgr progenitor, similar to the methodology adopted by \citet{limberg23}, who used the same $N$-body model as a reference.

This cross-match resulted in a sample of 2377 RRLs, which we refer to as RRLS-SGR-MODEL (Table~\ref{tab:rrls_sgr_model}). Among the RRLs in the RRLS-SGR-MODEL sample, 1926 stars (81\%) are also present in the RRLS-SGR-R22 sample, which was selected using completely independent criteria by \citet{Ramos2022}. This significant overlap confirms the robustness of our selection procedure, demonstrating that our cross-match with the model data effectively identifies RRLs associated with the Sgr stream. Figure~\ref{fig:4panel_model} shows the distributions of RRLs from the RRLS-SGR-MODEL sample alongside the simulated stars from \citet{Vasiliev2021}. Compared to the RRLS-SGR-R22 sample, the RRLS-SGR-MODEL sample is more sparsely populated in the region of the leading arm apocenter ($\Lambda \simeq 80\degr$, $D \simeq 60$~kpc), but it does sample with a handful of RRLs some arms of the stream that are not present in the RRLS-SGR-R22 sample.
%As expected, the RRLs from the RRLS-SGR-MODEL sample align very well with the simulated stars from \citet{Vasiliev2021}. 

\item We performed our own selection of RRLs belonging to the Sgr stream using a slightly modified version of the selection procedure introduced by \citet{Ibata2020}, which includes the following steps:

\begin{itemize}

\item From the clean sample of RRLs, we selected stars with distances from the Sun greater than 15~kpc and less than 150~kpc, and with absolute distance errors below 10~kpc. This resulted in a sample of 58,547 stars.

\item As shown in Fig.~\ref{fig:lambda_beta}, members of the Sgr stream have Sgr $B$ coordinate within the range [$-20^\circ$, $20^\circ$]. We therefore selected RRLs within this range, yielding a sample of 14,805 stars.

\item We transformed the Sgr coordinates used in our study, defined in the system introduced by \citet{Vasiliev2021}, into the system adopted by \citet{Ibata2020} in the following way: 
\begin{equation}
\begin{aligned}
    \Lambda_{\rm I20} = -\Lambda_{\rm V21} \\
    \mu_{\Lambda_{\rm I20}} = -\mu_{\Lambda_{\rm V21}} \\
    \mu_{B_{\rm I20}} = -\mu_{B_{\rm V21}}
\end{aligned}
\end{equation}

\item We then applied the equation proposed by \citet{Ibata2020} in the form:
\begin{equation}\label{eq:lambda_fit}
\mu_{\Lambda_{\rm I20}, fit}(\Lambda_{\rm I20}) = a_1\sin(a_2\Lambda_{\rm I20} + a_3) + a_4 + a_5\Lambda_{\rm I20} + a_6\Lambda_{\rm I20}^2,    
\end{equation}

where we slightly modified the coefficients $a_3$ and $a_4$\footnote{This modification of the \citet{Ibata2020} selection criteria was made to adjust the fit to the distribution of the Sgr stream as traced by the {\it Gaia} DR3 data, since the original formula by \citet{Ibata2020} was based on the {\it Gaia} DR2 data.} resulting in the following values: $a_1 = 1.1842$, $a_2 = -1.5639 \times \pi/180$ , $a_3 = -0.2$,
$a_4 = -2$, $a_5 = -8.0606 \times 10^{-4}$, and
$a_6 = 3.2441 \times 10^{-5}$. We then selected stars satisfying the criterion $|\mu_{\Lambda_{\rm I20}} -\mu_{\Lambda_{\rm I20}, fit}| < 0.6$ and obtained a sample of 6738 RRLs.

\item We then selected stars based on $\mu_B$, adopting the fitted function $\mu_{B_{\rm I20}, \text{fit}}(\Lambda_{\rm I20})$, which follows the same functional form as $\mu_{\Lambda_{\rm I20}, \text{fit}}(\Lambda_{\rm I20})$ in Eq.~\ref{eq:lambda_fit}, with the coefficients from \citet{Ibata2020}: $a_1 = -1.2360$, $a_2 = 1.0910 \times \pi/180$, $a_3 = 0.3633$, $a_4 = -1.3412$, $a_5 = 7.3022 \times 10^{-3}$, and $a_6 = -4.3315 \times 10^{-5}$. We retained stars satisfying the criterion $|\mu_{B_{\rm I20}} - \mu_{B_{\rm I20}, \text{fit}}| < 0.6$, resulting in a sample of 5271 RRLs.

The distribution of RRLs selected in this way is shown with red points in Fig.~\ref{fig:4panel_bellazzini}, overlaid on the present-day snapshot from the \citet{Vasiliev2021} model (black dots). Panel (a) shows their distribution in the Cartesian $Z$ versus $X$ plane; panel (b) presents the distribution of distances versus the Sgr $\Lambda$ coordinate for the same samples; while panels (c) and (d) display the {\it Gaia} DR3 proper motions, transformed into Sgr coordinates, $\mu_\Lambda$ and $\mu_B$, as functions of $\Lambda$. The Sgr coordinates follow the system defined by \citet{Vasiliev2021}. Figs.~\ref{fig:4panel_ramos} and \ref{fig:4panel_model}, which show the properties of the RRLS-SGR-R22 and RRLS-SGR-MODEL samples, respectively, are arranged in the same way.

The leading and trailing arms of the Sgr stream are clearly visible in Fig.~\ref{fig:4panel_bellazzini}, sampled with more stars and extending to larger distances from the main body along the stream than in the RRLS-SGR-R22 and RRLS-SGR-MODEL samples. However, this approach, as well as the previous ones, is unable to select the RRLs around the apocenter of the trailing arm identified by \citet{Sesar2017} and noted by \citet{bellaz20} in the region of the globular cluster NGC2419 \citep[see also][]{belok14,davies24}.
In an attempt to include these stars, we applied an additional selection step by choosing, from the clean sample of RRLs, stars with:

\begin{equation}\label{eq:far}
\begin{aligned}
&\Lambda_{\rm V21} > 130^\circ \\
-20^\circ &\leq B_{V21} \leq 20^\circ \\
60 &< D_\odot < 150~{\rm kpc} \\
&\sigma_{D_\odot} < 10~{\rm kpc} \\ 
0 &\leq \mu_{\Lambda_{\rm V21}} \leq 0.8~{\rm mas/yr} \\
0.1 &\leq \mu_{B_{\rm V21}} \leq 1~{\rm mas/yr}
\end{aligned}
\end{equation}

This selection allowed us to identify 25 additional RRLs as candidate members of the Sgr stream (shown with blue points in Fig.~\ref{fig:4panel_bellazzini}), located in the lower branch of the bifurcation occurring at the apocenter of the trailing arm, that is, the branch bending back toward the Galactic center. We will refer to this sparse sample, which provides an insight into the old population in this remote portion of the Sgr stream, as the "far arm". Together with the 5271 RRLs selected in the previous steps, they form a sample of 5296 RRLs, which we refer to as RRLS-SGR-SELECTION in the following analysis (Table~\ref{tab:rrls_sgr_selection}). This selection procedure allowed us to identify 3500 out of 3865 (90\%) RRLs from the RRLS-SGR-R22 sample as Sgr stream members, confirming the robustness of our method.

\end{itemize}

\end{enumerate}

Information on the three datasets selected in steps (1)–(3) is presented in Table~\ref{tab:subsets}. Fig.~\ref{fig:met_hist} shows that the three samples have very similar metallicity distributions and indistinguishable weighted mean metallicity values (see Table~\ref{tab:subsets}). The mean metallicity of RRLs in the Sgr stream, calculated as the mean value of metallicities from the three samples RRLS-SGR-R22, RRLS-SGR-SELECTION, and RRLS-SGR-MODEL used in this study, is [Fe/H]=$-1.62 \pm 0.01$~dex, in excellent agreement with \citet{Ramos2020}. Fig.~\ref{fig:4panel_distance} shows the distributions of stars simulated in the \citet{Vasiliev2021} model and RRLs from the RRLS-SGR-MODEL, RRLS-SGR-R22, and RRLS-SGR-SELECTION samples on the $\mu_\Lambda$ versus $\Lambda$ plane, color-coded by distances. There is good agreement between the model and all three datasets, confirming that the model successfully reproduces the main features of the Sgr stream as traced by RRLs. While the agreement with the RRLS-SGR-MODEL sample is expected, the consistency with the other two samples provides an independent validation. Additionally, Fig.~\ref{fig:4panel_distance} highlights the different properties and selection effects affecting each of the three samples.

Fig.~\ref{fig:H17_distance} shows the distribution of stars simulated in the \citet{Vasiliev2021} model, along with RRLs from the RRLS-SGR-R22 and RRLS-SGR-SELECTION samples, plotted in the distance versus $\Lambda$ plane. The squares represent the mean distances as a function of $\Lambda$ modeled by \citet{Hernitschek2017}, based on RRLs from the PS1 survey with distances derived using $PL$ relations in the visual and near-infrared PS1 bands \citep{Sesar2017b}. The figure demonstrates a good agreement between the distance estimates in this study and the independent results from \citet{Hernitschek2017}.

The RRLS-SGR-R22 sample is expected to have a high degree of purity; however, it is limited to $-140\degr \leq \Lambda \leq 150\degr$, and it also lacks stars in the range $5\degr \leq \Lambda \leq 30\degr$ to avoid contamination from stars in the Galactic Disk. Additionally, it does not include RRLs from the far arm of the Sgr stream. The RRLS-SGR-MODEL sample, while less complete and dependent on the \citet{Vasiliev2021} model, provides valuable information on the stripping times of the various portions of the stream. Finally, the RRLS-SGR-SELECTION sample is the most complete, including RRLs in regions much closer to the Galactic Disk than RRLS-SGR-R22, and from the far arm, but it also has a higher probability of containing non-members of the Sgr stream.

In the following analysis, we will primarily present results using RRLS-SGR-SELECTION as the reference sample. However, all results presented in the following sections were, when possible, also cross-checked with the other two samples. Although we do not explicitly report this in each case, the findings were consistently in good agreement. This approach strongly mitigates any potential dependency of the presented results on specific selection criteria or samples.

\section{Metallicity of the Sgr stream}\label{sec:met}

\subsection{Metallicity gradient in the Sgr stream}\label{sec:arms}

\begin{figure*}
\includegraphics[width=18cm]{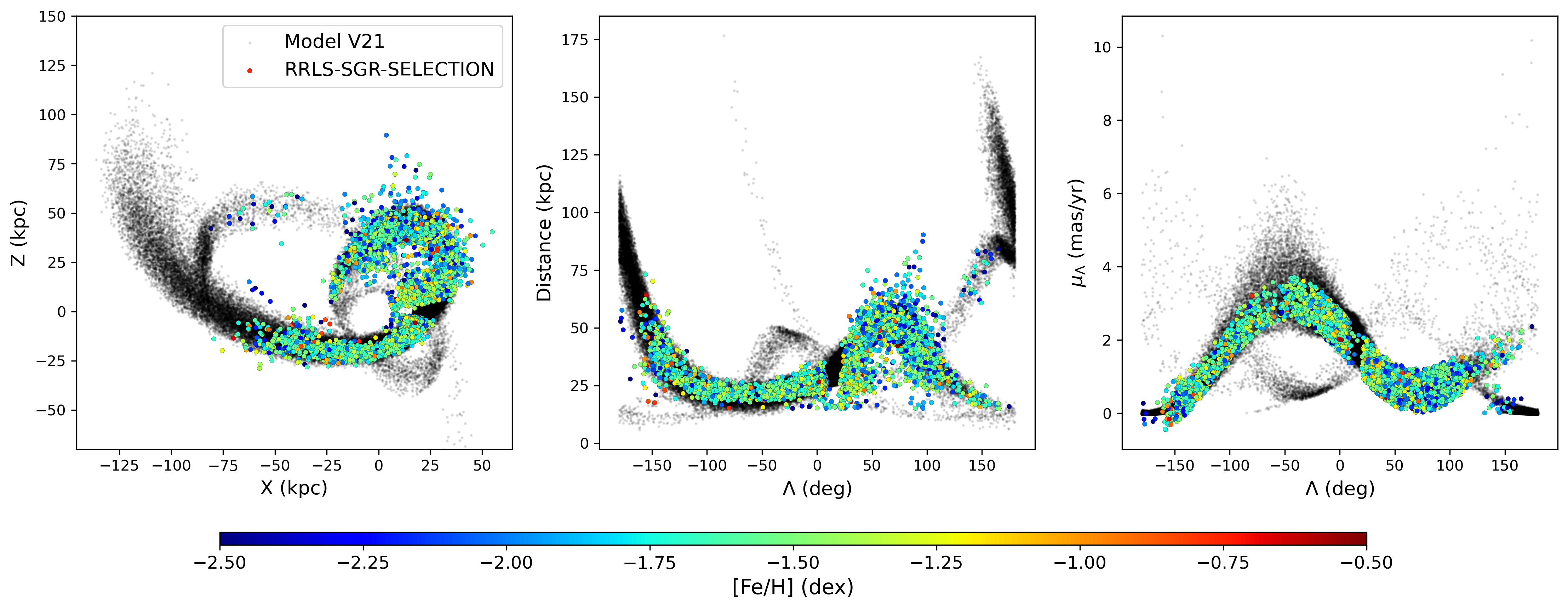}
\includegraphics[width=18cm]{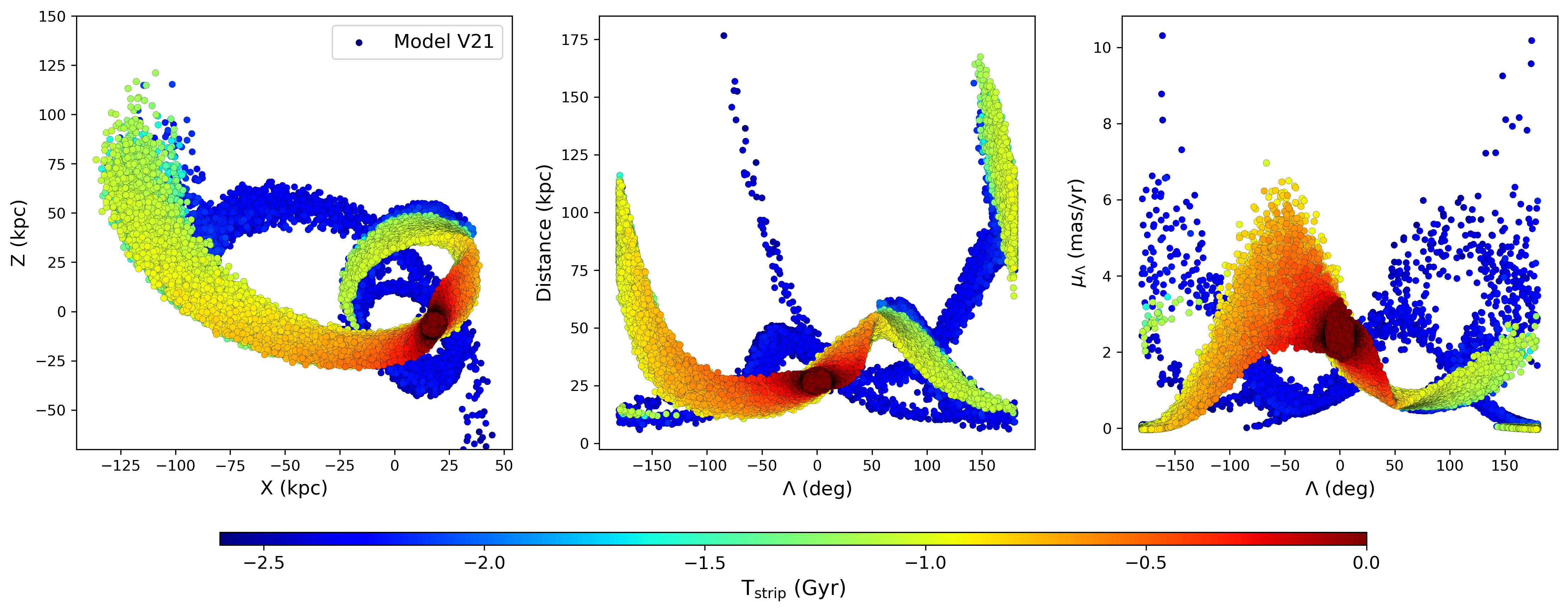}
\caption{{\it Upper panels}: Distributions of RRLs from the RRLS-SGR-SELECTION sample in the Cartesian $Z$ versus $X$ plane, the heliocentric distance versus the Sgr $\Lambda$ coordinate plane, and the proper motion $\mu_\Lambda$ versus $\Lambda$ plane, color-coded by metallicity.
{\it Lower panels}: The same distributions for the sample of simulated stars from \citet{Vasiliev2021}, color-coded by time of stripping from the Sgr progenitor.}\label{fig:6panels}
\end{figure*}

\begin{figure}
\includegraphics[width=\linewidth, trim=0 0 40 0]{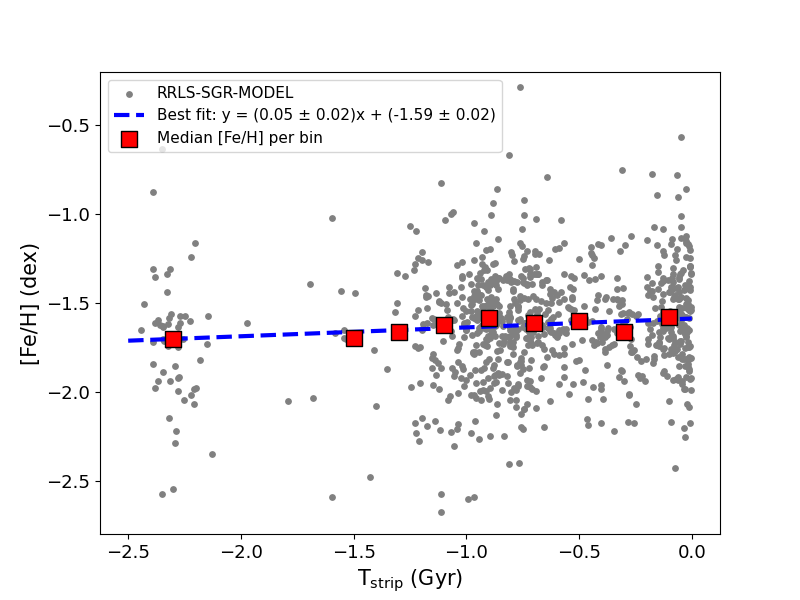}
\caption{Metallicity as a function of stripping time ($\rm T_{strip}$) for RRLs from the RRLS-SGR-MODEL sample with ${\rm T_{strip} < 0}$ (grey dots). The red squares indicate median metallicities computed within bins containing more than ten RRLs. The blue dashed line represents the best linear fit to the median metallicity values.} \label{fig:metallicity_tstrip}
\end{figure}

\begin{figure*}
\includegraphics[width=18cm]{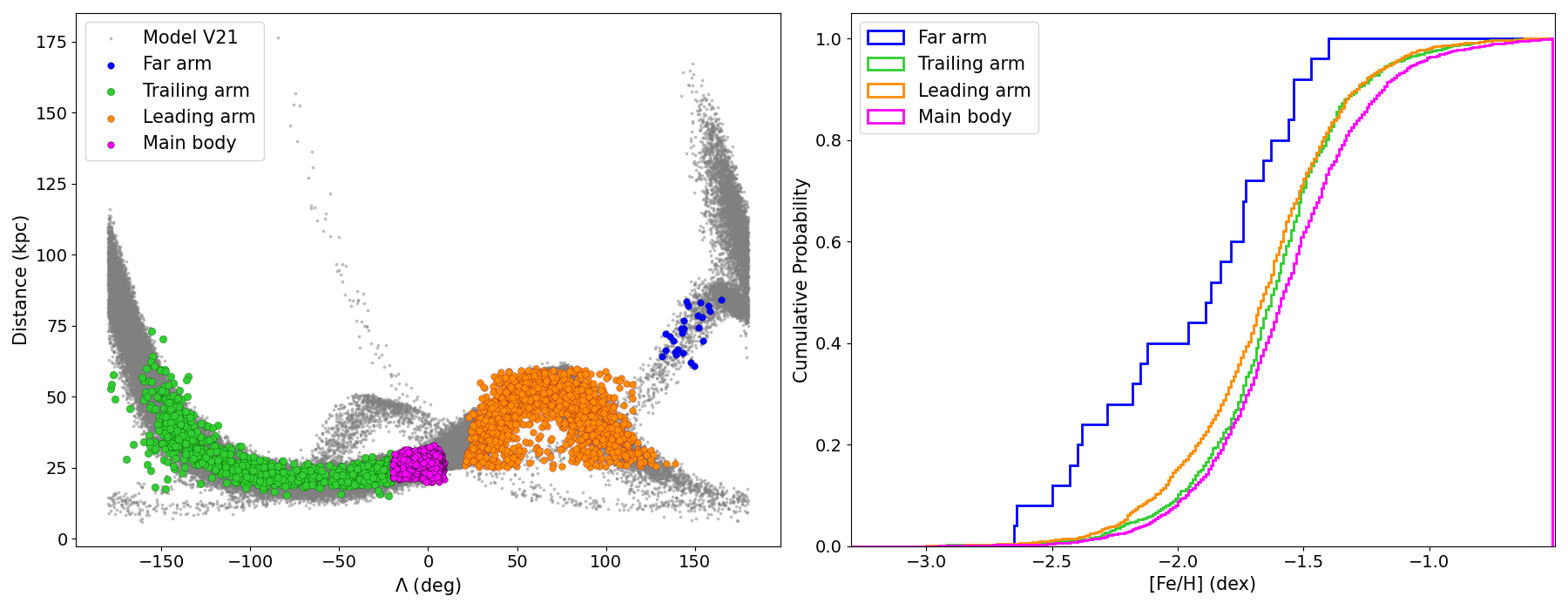}
\caption{{\it Left panel:} Distribution of RRLs from different Sgr stream substructures in the distance versus $\Lambda$ plane.
{\it Right panel:} Cumulative distribution functions of metallicity for RRLs in each substructure.}\label{fig:arms_division}
\end{figure*}

\begin{figure}
\includegraphics[width=\linewidth, trim=0 0 40 0]{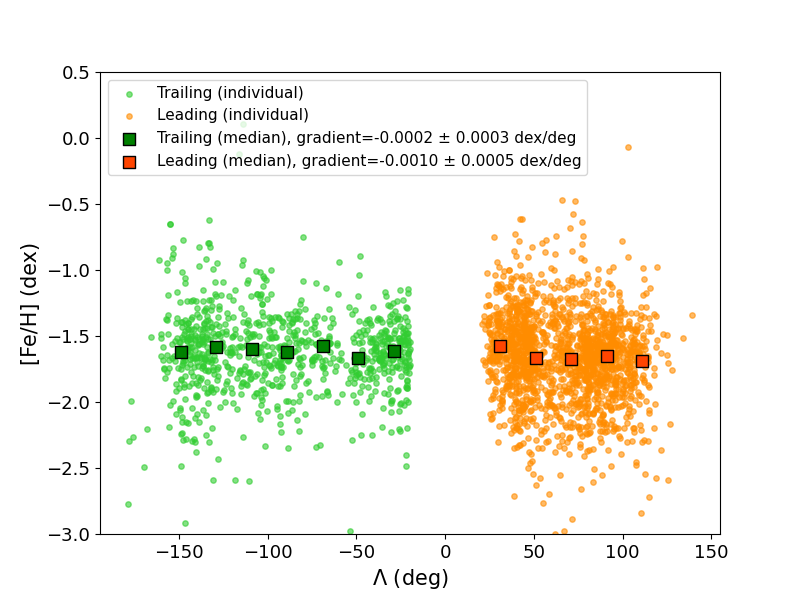}
\caption{Distribution of metallicity as a function of the Sgr  $\Lambda$ coordinate for RRLs in the trailing (green points) and leading (orange points) arms. Dark green and red squares represent median metallicities computed in bins containing more than 15 RRLs for the trailing and leading arms, respectively.}\label{fig:lambda_metallicity} 
\end{figure}

In this section, we study the metallicity distribution of RRLs in the Sgr stream. It is important to recall that RRLs do not trace the metallicity gradient of the entire population of the Sgr system, nor of its dominant component, which is likely younger than $\simeq 8$–10~Gyr and more metal-rich than [Fe/H]$\simeq -1.0$~dex \citep{deboer15}, especially in the main body and in the stream regions closest to it \citep{bellaz06a,minelli23}. On the other hand, since RRLs are mainly old stars (age $\geq 10$~Gyr), and the vast majority of them have [Fe/H] $\le -1.0$~dex, they allow us to study the metallicity distribution of the oldest and, consequently, most metal-poor stellar populations, giving a glance to the dynamical evolutionary history of stars that belonged to the Sgr progenitor before its infall into the MW halo.

The upper panels of Fig.~\ref{fig:6panels} show the distributions of RRLs from the RRLS-SGR-SELECTION sample, color-coded by metallicity. For comparison, the lower panels of Fig.~\ref{fig:6panels} display the distributions of simulated stars from the \citet{Vasiliev2021} model, color-coded by stripping time from the Sgr progenitor. 
%As shown, there is a concentration of metal-rich RRLs in the Sgr core, while a group of slightly more metal-poor RRLs is located in the outskirts of the Sgr progenitor. 
The leading and far arms of the Sgr stream, which, according to the model of \citet{Vasiliev2021}, contain stars stripped in earlier epochs, appear visually more metal-poor. This suggests that the metallicity gradient in the Sgr progenitor—with metal-rich stars concentrated in the center and metal-poor stars in the halo—was already established at the time of RRLs formation ($\sim$10 Gyr ago).

To study the correlation between the epoch at which RRLs were stripped from the Sgr dSph and their metallicities, in Fig.~\ref{fig:metallicity_tstrip}, we plot the metallicities of RRLs in the RRLS-SGR-MODEL sample as a function of stripping time, based on the model of \citet{Vasiliev2021}. Red squares indicate the median metallicities in bins containing more than ten RRLs. Only stars with stripping times less than zero, which have already left the sphere with a radius of 5 kpc around the Sgr progenitor \citep{Vasiliev2021}, are shown. 
We found a weak metallicity gradient of $0.05\pm0.02$ dex/Gyr, determined by performing a linear least-squares fit to the median metallicity values in bins of stripping time. The fit yields a $p$-value of 0.026, representing the probability of obtaining the observed gradient under the null hypothesis (i.e., if the true underlying gradient were zero). Since this value is below the conventional 0.05 threshold, we conclude that the metallicity gradient is statistically significant at the 95\% confidence level. This result suggests that a metallicity gradient was present in the Sgr progenitor already at very early epochs, with less metal-poor stars concentrated toward the center and more metal-poor stars distributed in the halo.

To analyze the metallicity gradient in the Sgr main body and stream in more detail, we divide the RRLs in the RRLS-SGR-SELECTION sample into four subsets as follows:

\begin{itemize}
\item Leading arm: $10^\circ \leq \Lambda_{\rm V21} \leq 180^\circ$, $25 \leq D_\odot \leq  60$ kpc;
\item Trailing arm: $-180^\circ \leq \Lambda_{\rm V21} < -20^\circ$, $15 \leq D_\odot \leq  75$ kpc;
\item Main body: $-20^\circ \leq \Lambda_{\rm V21} < 10^\circ$, $20 \leq D_\odot \leq  33$ kpc;
\item Far arm: selected as in Eq.~\ref{eq:far}. 
\end{itemize}

The left panel of Fig.~\ref{fig:arms_division} shows the distribution of Sgr substructures in the distance versus $\Lambda$ plane, while the right panel presents the cumulative distribution functions of RRL metallicities for different substructures. Table~ \ref{tab:structures} summarizes the number of RRLs in each substructure along with their mean metallicities. These results agree well with the mean metallicities for the main body ($-1.60$~dex), leading arm ($-1.71$~dex), and trailing arm ($-1.65$~dex) reported by \citet{Sun2025}, who used photometric metallicities of RRLs derived by \citet{Li2023}. Table~\ref{tab:KS} shows the results of the two-sample Kolmogorov–Smirnov (KS) test comparing the metallicity distributions of RRLs in different substructures. All $p$-values are significantly below the canonical 0.05 threshold, indicating that the metallicity distributions in different substructures, are statistically different. As evident from Fig.~\ref{fig:arms_division} and Table~\ref{tab:structures}, the main body is the most metal-rich substructure of the Sgr stream, followed by the slightly more metal-poor trailing arm, then the leading arm, and finally the far arm, which is significantly more metal-poor. 
This trend is in agreement with the findings of \citet{Yang2019}, who analysed the kinematic and chemical properties of $\sim$3000 Sgr members, including K-giants, M-giants, and BHB stars. All three tracers indicate that the trailing arm is, on average, more metal-rich than the leading arm, while the K-giants also show that the far arm is the most metal-poor substructure. \citet{Ramos2022} and \citet{Sun2025} found a similar trend in the metallicity distributions of the main body, leading, and trailing arms, as traced by RRLs.

We find that the difference between the mean metallicity of RRLs in the leading and trailing arms (0.05~dex) is significantly smaller than the difference observed in these substructures using intermediate-age stars, which is, on average, 0.3~dex (\citealt{Yang2019}; \citealt{Hayes2020}; \citealt{Ramos2022}). This is likely because the metallicity gradient of intermediate-age stars was influenced by enrichment that occurred due to star formation episodes triggered by Sgr accretion onto the MW. The stars in the trailing arm were stripped from the Sgr progenitor at a later time, meaning they had more time to undergo chemical enrichment within the progenitor galaxy before being pulled away. The stars in the leading arm were stripped earlier, missing later enrichment events and thus exhibiting lower metallicity. In contrast, RRLs reflect the metallicity gradient of the ancient, metal-poor population, which was already established before the Sgr accretion onto the MW and is therefore much less pronounced.

%However, we observe a significant difference (0.4~dex) between the mean metallicities of the main body and the far arm (Table~\ref{tab:structures}). The reason we see this difference between the far arm and the main body, but not between the trailing and leading arms, is likely due to the narrow metallicity distribution of RRLs, which naturally results in a smaller observed difference between substructures. Additionally, the photometric metallicities of RRLs have larger uncertainties compared to spectroscopic measurements, further diluting the observed gradient.

In Fig.~\ref{fig:lambda_metallicity}, we present the distribution of metallicity as a function of the Sgr $\Lambda$ coordinate for the trailing and leading arms. The green and red squares represent the median metallicities of the trailing and leading arms, respectively, calculated for bins containing more than 15 RRLs. We find negligible metallicity gradients of $(-0.2\pm0.3)\times 10^{-3}$ dex/deg in the trailing arm and of $(-1.0\pm0.5)\times 10^{-3}$ dex/deg in the leading arm in agreement with previous studies \citep{Ramos2020, Sun2025}.

\begin{table}
\caption{Weighted mean metallicity of RRLs in different substructures of the Sgr stream}\label{tab:structures}
\centering
\begin{tabular}{lccc}
\hline \hline
 Name & N & [Fe/H] range & Mean [Fe/H] \\
   &  (RRLs) & (dex) & (dex) \\
 \hline
Main body & 2445 & [-2.94, +0.62] & $-1.58 \pm 0.31$ \\
Trailing arm & 865 & [-2.98, +0.10] & $-1.64 \pm 0.28$ \\
Leading arm & 1690 & [-3.00, -0.07] & $-1.69 \pm 0.31$ \\
Far arm & 25 & [-2.64, -1.40] & $-1.98 \pm 0.37$ \\
\hline
\end{tabular}
\end{table}

\begin{table}
\caption{Results of the two-sample KS test comparing the metallicity distributions of RRLs in different substructures}\label{tab:KS}
\centering
\begin{tabular}{lc}
\hline \hline
& $p$-value \\
 \hline
Far arm - Main body &  $9\times 10^{-5}$ \\
Far arm - Trailing arm  & $4\times 10^{-4}$ \\
Far arm - Leading arm &  $5\times 10^{-3}$ \\
Main body - Trailing arm & $3\times 10^{-6}$\\
Main body - Leading arm & $1 \times 10^{-15}$\\
Trailing arm - Leading arm & $1 \times 10^{-3}$ \\
\hline
\end{tabular}
\end{table}

\subsection{Bifurcations}

\begin{figure}
\includegraphics[width=\linewidth, trim=0 0 0 0]{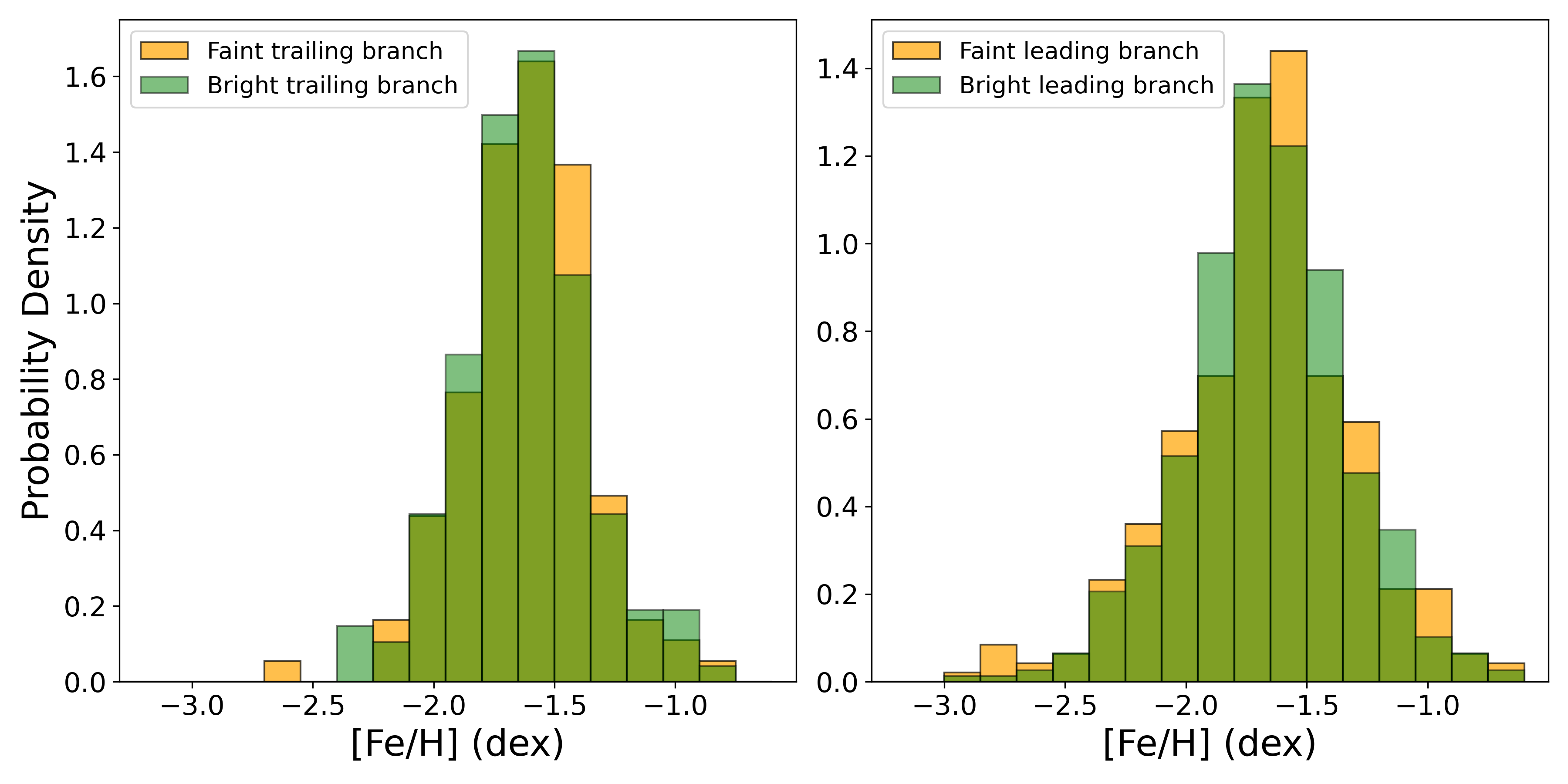}
\caption{Metallicity distributions of RRLs in the faint (orange) and bright (green) branches of the trailing ({\it left panel}) and leading arms ({\it right panel}) of the Sgr stream.}\label{fig:bifur} 
\end{figure}

\begin{table}
\caption{Weighted mean metallicity of RRLs in the faint and bright branches of the Sgr stream}\label{tab:bifur}
\centering
\begin{tabular}{lccc}
\hline \hline
 Name & N &  Mean [Fe/H] \\
   &  (RRLs) & (dex) \\
 \hline
Faint leading branch &  316  & $-1.65 \pm 0.40$ \\
Bright leading branch &  519  &  $-1.64 \pm 0.35$ \\
Faint trailing branch &  122  &  $-1.61 \pm 0.28$ \\
Bright trailing branch &  316  &  $-1.62 \pm 0.28$ \\
\hline
\end{tabular}
\end{table}

\citet{Belokurov2006} discovered the two branches in the leading arm of the Sgr stream, commonly referred to as a bifurcation. Later, \citet{Koposov2012} and \citet{Navarrete2017} identified a similar bifurcation, comprising faint and bright branches, in the trailing arm. Although several models have been proposed to explain this feature (e.g., \citealt{Penarrubia2011}), its origin remains a subject of debate.
Recently, \citet{Ramos2022} suggested that the bifurcation arises from a misaligned overlap of material ejected at the antepenultimate pericenter (faint branch) with stars stripped at the penultimate pericenter (bright branch). It was also found that the faint branch is more metal-poor than the bright branch (\citealt{Koposov2012}; \citealt{Ramos2022}).

In this study, we used RRLs from the RRLS-SGR-R22 sample to analyze the metallicity distribution of the faint and bright branches in both the leading and trailing arms. \citet{Ramos2022} assigned a probability to each star in their sample indicating its likelihood of belonging to either the faint (\texttt{ProbFaint}) or the bright branch (\texttt{ProbBright}). For our analysis, we applied selection criteria consistent with those adopted in Section~\ref{sec:arms} to identify the faint and bright branches of the leading and trailing arms:

\begin{itemize} 
\item Faint leading branch: $10^\circ \leq \Lambda_{\rm V21} \leq 180^\circ$, \texttt{ProbFaint}~>~0.2, \texttt{ProbFaint }~>~\texttt{ProbBright}; 
\item Bright leading branch: $10^\circ \leq \Lambda_{\rm V21} \leq 180^\circ$, \texttt{ProbBright}~>~0.2, \texttt{ProbFaint}~<~\texttt{ProbBright}; 
\item Faint trailing branch: $-180^\circ \leq \Lambda_{\rm V21} < -20^\circ$, 
\texttt{ProbFaint}~>~0.2, \texttt{ProbFaint}~>~\texttt{ProbBright}; 
\item Bright trailing branch: $-180^\circ \leq \Lambda_{\rm V21} < -20^\circ$, \texttt{ProbBright}~>~0.2, \texttt{ProbFaint}~<~\texttt{ProbBright}; 
\end{itemize}

Table~\ref{tab:bifur} presents the number of RRLs in each of these substructures along with their mean metallicity, while Fig.~\ref{fig:bifur} shows the metallicity distributions. The faint branch of the leading arm is only 0.01~dex more metal-poor than the bright branch, whereas the faint branch of the trailing arm is 0.01~dex more metal-rich than the bright branch.
Since the 0.01~dex difference is not statistically significant, we conclude that the metallicity difference between the faint and bright branches of the Sgr stream is not confirmed based on the RRLs in our sample. This result is consistent with the findings of \citet{Ramos2020}.

\subsection{Main body}

\begin{figure}
\includegraphics[width=\linewidth, trim=0 0 40 0]{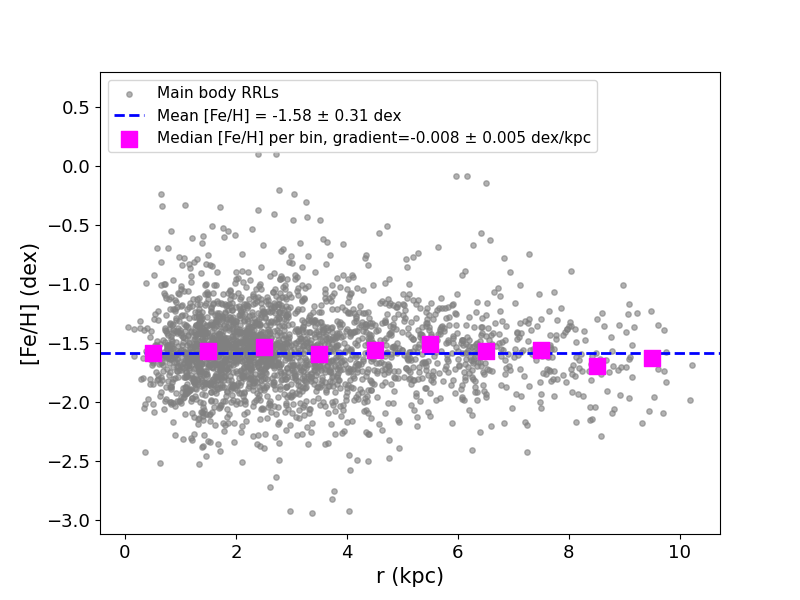}
\caption{Metallicity distribution as a function of radial distance from the center of the Sgr core for RRLs in the RRLS-SGR-SELECTION  sample belonging to the main body. The blue line represents the weighted mean metallicity of RRLs in the main body, while the magenta squares indicate the median metallicities calculated for each bin containing more than 10 RRLs.}\label{fig:main_body} 
\end{figure}

In Section~\ref{sec:arms}, we identified 2445 RRLs from the RRLS-SGR-SELECTION sample as belonging to the main body of the Sgr stream. For each RRL in this sample, we calculated the radial distance from the center of the Sgr core using the Euclidean distance between the Galactocentric Cartesian coordinates of the Sgr center (\{17.9, 2.6, -6.6\},  \citealt{Vasiliev2021}) and the coordinates of the star, as derived in Section~\ref{sec:gaia}.
Fig.~\ref{fig:main_body} illustrates the metallicity of RRLs in the main body as a function of radius. The blue dashed line represents the weighted mean metallicity of RRLs in the main body, ${\rm [Fe/H]} = -1.58 \pm 0.31$~dex. The magenta squares indicate the median metallicities calculated for each bin containing more than 10 RRLs.
We find a metallicity gradient of $-0.008\pm0.005$ dex/kpc, with more metal-rich RRLs located closer to the center of the Sgr progenitor. However, this gradient is not statistically significant ($p$-value = 0.198), indicating that the observed trend is consistent with no real correlation between metallicity and radial distance.

\subsection{Far arm}

We analyzed 25 RRLs belonging to the far arm of the Sgr stream, selected as described in Eq.~\ref{eq:far}. The weighted mean metallicity of RRLs in the far arm is ${\rm [Fe/H]} = -1.98 \pm 0.37$~dex, which is approximately 0.3~dex lower than the mean metallicity of the RRLs in the leading arm. This makes the far arm the most metal-poor structure of the Sgr stream. This finding aligns with the expectation that more metal-poor RRLs were originally located in the outskirts of the Sgr progenitor and, hence, were stripped during the earlier stages of Sgr accretion.

\begin{figure}
\includegraphics[width=\linewidth, trim=0 0 0 0]{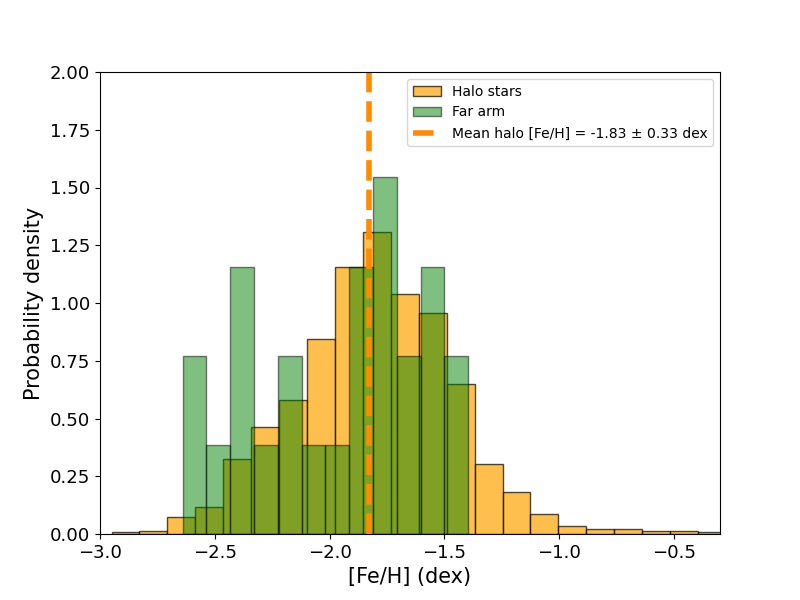}
\caption{Metallicity distributions of RRLs in the far arm of the Sgr stream (green bins) and in the MW halo at distances between 61 and 84~kpc (orange bins). The dark orange dashed line indicates the weighted mean metallicity of the MW halo sample.}\label{fig:far_arm_hist} 
\end{figure}

Fig.~\ref{fig:far_arm_hist} presents the histogram (green bins) of the metallicity distribution of RRLs in the far arm. Two distinct peaks are visible, approximately at ${\rm [Fe/H]} = -2.4$~dex and ${\rm [Fe/H]} = -1.7$~dex. This bimodal distribution could be due either to contamination from RRLs in the MW field or to the possibility that RRLs in the far arm were stripped from the Sgr progenitor at different epochs during the accretion process. We compare this distribution with the metallicity distribution of RRLs in the MW halo located at distances corresponding to those of the Sgr far arm RRLs (61–84~kpc). We excluded from this sample RRLs located in the Sgr far arm and within $10^\circ$ of the LMC and SMC centres, resulting in a sample of 1118 RRLs. Their weighted mean metallicity is ${\rm [Fe/H]} = -1.83\pm0.33$~dex. The metallicity distribution of this halo sample is shown with orange bins in Fig.~\ref{fig:far_arm_hist}. Although one of the peaks in the far arm distribution is close to the peak of the halo sample, the two distributions appear visually distinct. 
A two-sample KS test between the MW halo stars and the metal-rich ([Fe/H] > –2~dex) subset of the Sgr far arm yields a $p$-value of 0.067, indicating that the two samples could be drawn from the same distribution. In contrast, the KS test between the halo stars and the metal-poor ([Fe/H] < –2 dex) subset of the Sgr far arm returns a $p$-value of $10^{-7}$, confirming that these two datasets are statistically different.
However, the current sample size of far arm RRLs is too small to draw firm conclusions about the origin of its bimodal metallicity distribution.

\section{Conclusions}\label{sec:summ}

In this study, we analyzed the Sgr stream as traced by RRLs from the {\it Gaia} DR3 catalogue \citep{Clementini2023}. Photometric metallicities for these RRLs were computed by \citet{Muraveva2025}, and accurate distances were determined using the reddening-free $PWZ$ relation from \citet{Garofalo2022}. The new metallicity and distance estimates enable a more precise characterization of the Sgr stream’s structure and metallicity gradient. Unlike young and intermediate-age stars, RRLs are ideal tracers of the metallicity gradient in the Sgr progenitor as it existed before the accretion of the Sgr dSph by the MW, as they remain unaffected by the chemical enrichment from recent star formation episodes that occurred following the accretion.

We found good agreement between the distances and kinematic distribution of the RRLs in our sample and those of the stars simulated in the model of \citet{Vasiliev2021}, which accounts for the gravitational potential of the MW and the LMC. Using the simulated stars from the \citet{Vasiliev2021} model, we were able to calculate the metallicity gradient as a function of the time of stripping of the stars from the Sgr progenitor. We find a mild, but statistically significant metallicity gradient of $0.05\pm0.02$ dex/Gyr, with more metal-poor stars being stripped at earlier epochs. This result confirms the existence of a metallicity gradient in the Sgr progenitor at the epoch of formation of RRLs, with metal-rich stars concentrated toward the center and metal-poor stars distributed in the outer regions of the Sgr progenitor.

We found that the main body is the most metal-rich substructure of the Sgr stream ([Fe/H]=$-1.58\pm0.31$~dex), followed by the slightly more metal-poor trailing arm ([Fe/H]=$-1.64\pm0.28$~dex), then the leading arm ([Fe/H]=$-1.69\pm0.31$~dex), and finally the far arm ([Fe/H]=$-1.98\pm0.37$~dex), which is significantly more metal-poor. Additionally, we found almost negligible metallicity gradients as a function of $\Lambda$ with values of $(-0.2\pm0.3)\times 10^{-3}$ dex/deg in the trailing arm and of $(-1.0\pm0.5)\times 10^{-3}$ dex/deg in the leading arm. These results are in agreement with previous studies \citep{Ramos2020, Sun2025}. We also analyzed the bifurcation in the trailing and leading arms \citep{Belokurov2006, Koposov2012}. We found that the difference in mean metallicities between the faint and bright branches is approximately ~0.01 dex, which is not statistically significant. We conclude that the metallicity difference between the faint and bright branches of the Sgr stream is not confirmed based on the RRLs in our sample. This result is consistent with the findings of \citet{Ramos2020}.

To the best of our knowledge, in this study we analyze for the first time the metallicity gradient as a function of radial distance in the Sgr main body as traced by RRLs. We found a weak metallicity gradient of $-0.008\pm0.005$~dex/kpc, with more metal-rich RRLs located closer to the center of the Sgr progenitor. However, the observed gradient is not statistically significant, indicating that the true gradient may be consistent with zero. We also analyzed 25 RRLs in the far arm of the Sgr stream and found a bimodal distribution in metallicity, with peaks at [Fe/H]=$-$2.4~dex and [Fe/H]=$-$1.7~dex. This bimodal distribution could be due to contamination of the far arm sample by MW RRLs. Alternatively, the bimodality might reflect different epochs of stripping of RRLs in the far arm from the Sgr progenitor. However, our sample of RRLs in the far arm is not large enough to draw robust conclusions about the origin of this two-peaked metallicity distribution.

 The Sgr stream is one of the most significant contributors to the MW’s stellar halo. The mean metallicity of RRLs in the Sgr stream, calculated as the mean value of metallicities from the three samples used in this study, is [Fe/H]=$-1.62\pm 0.01$~dex. This value is in good agreement but slightly more metal-poor than the median metallicity of the MW halo RRLs [Fe/H]=$-1.55\pm 0.01$~dex \citep{Crestani2021}. At the same time, the Sgr stream’s RRLs are less metal-poor than those in some ultra-faint streams, such as the Orphan stream, which has a mean metallicity of [Fe/H]=$-1.80\pm0.06$~dex \citep{Prudil2021}. This reflects Sgr’s larger progenitor mass ($4.3 \times 10^7~{\rm M_\odot}$, \citealt{Pace2024}) and higher overall metallicity (see the discussion on the mass-metallicity relation of galaxies as traced with RRLs in \citealt{Bellazzini2025}). A more extensive study of the structure and metallicity gradient of the Sgr stream, as traced with RRLs, will be made possible with the arrival of the {\it Gaia} Data Release 4 (DR4), currently foreseen for the second half of 2026.

\begin{acknowledgements}

Support to this study has been provided by INAF Mini-Grant (PI: Tatiana Muraveva), by the Agenzia Spaziale Italiana (ASI) through contract ASI 2018-24-HH.0 and its Addendum 2018-24-HH.1-2022, and by Premiale 2015, MIning The Cosmos - Big Data and Innovative Italian Technology for Frontiers Astrophysics and Cosmology (MITiC; P.I.B.Garilli). MB acknowledge the support to this study by the PRIN INAF Mini Grant 2023 (Ob.Fu. 1.05.23.04.02 – CUP C33C23000960005) CHAM – Chemo-dynamics of the Accreted Halo of the Milky Way (P.I.: M. Bellazzini). MB acknowledges the financial support by the project LEGO – Reconstructing the building blocks of the Galaxy by chemical tagging (P.I. A. Mucciarelli), granted by the Italian MUR through contract PRIN 2022LLP8TK\_001. 
This work uses data from the European Space Agency mission {\it Gaia} (https://www.cosmos.esa.int/gaia), processed by the {\it Gaia} Data Processing and Analysis Consortium (DPAC; https: //www.cosmos.esa.int/web/gaia/dpac/consortium). Funding for the DPAC has been provided by national institutions, in particular the institutions participating in the {\it Gaia} Multilateral Agreement. 

\end{acknowledgements}

\bibliographystyle{aa}
\bibliography{bibl}

\begin{thebibliography}{80}
\expandafter\ifx\csname natexlab\endcsname\relax\def\natexlab#1{#1}\fi

\bibitem[{{Antoja} {et~al.}(2020){Antoja}, {Ramos}, {Mateu}, {Helmi}, {Anders}, {Jordi}, \& {Carballo-Bello}}]{Antoja2020}
{Antoja}, T., {Ramos}, P., {Mateu}, C., {et~al.} 2020, \aap, 635, L3

\bibitem[{{Bayer} {et~al.}(2025){Bayer}, {Starkenburg}, {Thomas}, {Martin}, {Helmi}, {Bystr{\"o}m}, {de Boer}, {Fern{\'a}ndez Alvar}, {Gwyn}, {Ibata}, {Jablonka}, {Kordopatis}, {Matsuno}, {McConnachie}, {Medina}, {S{\'a}nchez-Janssen}, \& {Sestito}}]{Bayer2025}
{Bayer}, M., {Starkenburg}, E., {Thomas}, G.~F., {et~al.} 2025, arXiv e-prints, arXiv:2502.17319

\bibitem[{{Bellazzini} {et~al.}(2006{\natexlab{a}}){Bellazzini}, {Correnti}, {Ferraro}, {Monaco}, \& {Montegriffo}}]{bellaz06a}
{Bellazzini}, M., {Correnti}, M., {Ferraro}, F.~R., {Monaco}, L., \& {Montegriffo}, P. 2006{\natexlab{a}}, \aap, 446, L1

\bibitem[{{Bellazzini} {et~al.}(2020){Bellazzini}, {Ibata}, {Malhan}, {Martin}, {Famaey}, \& {Thomas}}]{bellaz20}
{Bellazzini}, M., {Ibata}, R., {Malhan}, K., {et~al.} 2020, \aap, 636, A107

\bibitem[{{Bellazzini} {et~al.}(2025){Bellazzini}, {Muraveva}, \& {Garofalo}}]{Bellazzini2025}
{Bellazzini}, M., {Muraveva}, T., \& {Garofalo}, A. 2025, A\&A Letters [\eprint[arXiv]{2503.04925}], in press

\bibitem[{{Bellazzini} {et~al.}(2006{\natexlab{b}}){Bellazzini}, {Newberg}, {Correnti}, {Ferraro}, \& {Monaco}}]{bellaz06}
{Bellazzini}, M., {Newberg}, H.~J., {Correnti}, M., {Ferraro}, F.~R., \& {Monaco}, L. 2006{\natexlab{b}}, \aap, 457, L21

\bibitem[{{Belokurov} {et~al.}(2014){Belokurov}, {Koposov}, {Evans}, {Pe{\~n}arrubia}, {Irwin}, {Smith}, {Lewis}, {Gieles}, {Wilkinson}, {Gilmore}, {Olszewski}, \& {Niederste-Ostholt}}]{belok14}
{Belokurov}, V., {Koposov}, S.~E., {Evans}, N.~W., {et~al.} 2014, \mnras, 437, 116

\bibitem[{{Belokurov} {et~al.}(2006){Belokurov}, {Zucker}, {Evans}, {Gilmore}, {Vidrih}, {Bramich}, {Newberg}, {Wyse}, {Irwin}, {Fellhauer}, {Hewett}, {Walton}, {Wilkinson}, {Cole}, {Yanny}, {Rockosi}, {Beers}, {Bell}, {Brinkmann}, {Ivezi{\'c}}, \& {Lupton}}]{Belokurov2006}
{Belokurov}, V., {Zucker}, D.~B., {Evans}, N.~W., {et~al.} 2006, \apjl, 642, L137

\bibitem[{{Bobrick} {et~al.}(2024){Bobrick}, {Iorio}, {Belokurov}, {Vos}, {Vu{\v{c}}kovi{\'c}}, \& {Giacobbo}}]{Bobrick2024}
{Bobrick}, A., {Iorio}, G., {Belokurov}, V., {et~al.} 2024, \mnras, 527, 12196

\bibitem[{{Bono} {et~al.}(2003){Bono}, {Caputo}, {Castellani}, {Marconi}, {Storm}, \& {Degl'Innocenti}}]{Bono2003}
{Bono}, G., {Caputo}, F., {Castellani}, V., {et~al.} 2003, \mnras, 344, 1097

\bibitem[{{Carretta} {et~al.}(2009){Carretta}, {Bragaglia}, {Gratton}, {D'Orazi}, \& {Lucatello}}]{Carretta2009}
{Carretta}, E., {Bragaglia}, A., {Gratton}, R., {D'Orazi}, V., \& {Lucatello}, S. 2009, \aap, 508, 695

\bibitem[{{Catelan}(2004)}]{catelan04}
{Catelan}, M. 2004, in Astronomical Society of the Pacific Conference Series, Vol. 310, IAU Colloq. 193: Variable Stars in the Local Group, ed. D.~W. {Kurtz} \& K.~R. {Pollard}, 113

\bibitem[{{Chou} {et~al.}(2007){Chou}, {Majewski}, {Cunha}, {Smith}, {Patterson}, {Mart{\'\i}nez-Delgado}, {Law}, {Crane}, {Mu{\~n}oz}, {Garcia L{\'o}pez}, {Geisler}, \& {Skrutskie}}]{chou07}
{Chou}, M.-Y., {Majewski}, S.~R., {Cunha}, K., {et~al.} 2007, \apj, 670, 346

\bibitem[{{Clementini} {et~al.}(2003){Clementini}, {Gratton}, {Bragaglia}, {Carretta}, {Di Fabrizio}, \& {Maio}}]{Clementini2003}
{Clementini}, G., {Gratton}, R., {Bragaglia}, A., {et~al.} 2003, \aj, 125, 1309

\bibitem[{{Clementini} {et~al.}(2023){Clementini}, {Ripepi}, {Garofalo}, {Molinaro}, {Muraveva}, {Leccia}, {Rimoldini}, {Holl}, {Jevardat de Fombelle}, {Sartoretti}, {Marchal}, {Audard}, {Nienartowicz}, {Andrae}, {Marconi}, {Szabados}, {Evans}, {Lecoeur-Taibi}, {Mowlavi}, {Musella}, \& {Eyer}}]{Clementini2023}
{Clementini}, G., {Ripepi}, V., {Garofalo}, A., {et~al.} 2023, \aap, 674, A18

\bibitem[{{Clementini} {et~al.}(2019){Clementini}, {Ripepi}, {Molinaro}, {Garofalo}, {Muraveva}, {Rimoldini}, {Guy}, {Jevardat de Fombelle}, {Nienartowicz}, {Marchal}, {Audard}, {Holl}, {Leccia}, {Marconi}, {Musella}, {Mowlavi}, {Lecoeur-Taibi}, {Eyer}, {De Ridder}, {Regibo}, {Sarro}, {Szabados}, {Evans}, \& {Riello}}]{Clementini2019}
{Clementini}, G., {Ripepi}, V., {Molinaro}, R., {et~al.} 2019, \aap, 622, A60

\bibitem[{{Crestani} {et~al.}(2021){Crestani}, {Fabrizio}, {Braga}, {Sneden}, {Preston}, {Ferraro}, {Iannicola}, {Bono}, {Alves-Brito}, {Nonino}, {D'Orazi}, {Inno}, {Monelli}, {Storm}, {Altavilla}, {Chaboyer}, {Dall'Ora}, {Fiorentino}, {Gilligan}, {Grebel}, {Lala}, {Lemasle}, {Marengo}, {Marinoni}, {Marrese}, {Mart{\'\i}nez-V{\'a}zquez}, {Matsunaga}, {Mullen}, {Neeley}, {Prudil}, {da Silva}, {Stetson}, {Th{\'e}venin}, {Valenti}, {Walker}, \& {Zoccali}}]{Crestani2021}
{Crestani}, J., {Fabrizio}, M., {Braga}, V.~F., {et~al.} 2021, \apj, 908, 20

\bibitem[{{Cui} {et~al.}(2012){Cui}, {Zhao}, {Chu}, {Li}, {Li}, {Zhang}, {Su}, {Yao}, {Wang}, {Xing}, {Li}, {Zhu}, {Wang}, {Gu}, {Luo}, {Xu}, {Zhang}, {Liu}, {Zhang}, {Yang}, {Cao}, {Chen}, {Chen}, {Chen}, {Chen}, {Chu}, {Feng}, {Gong}, {Hou}, {Hu}, {Hu}, {Hu}, {Jia}, {Jiang}, {Jiang}, {Jiang}, {Jin}, {Li}, {Li}, {Li}, {Liu}, {Liu}, {Lu}, {Mao}, {Men}, {Qi}, {Qi}, {Shi}, {Tang}, {Tao}, {Wang}, {Wang}, {Wang}, {Wang}, {Wang}, {Wang}, {Wang}, {Wang}, {Wang}, {Wang}, {Wang}, {Wang}, {Xu}, {Xu}, {Yang}, {Yu}, {Yuan}, {Yuan}, {Zhai}, {Zhang}, {Zhang}, {Zhang}, {Zhao}, {Zhou}, {Zhou}, {Zhu}, \& {Zou}}]{Cui2012}
{Cui}, X.-Q., {Zhao}, Y.-H., {Chu}, Y.-Q., {et~al.} 2012, Research in Astronomy and Astrophysics, 12, 1197

\bibitem[{{Davies} {et~al.}(2024){Davies}, {Belokurov}, {Monty}, \& {Evans}}]{davies24}
{Davies}, E.~Y., {Belokurov}, V., {Monty}, S., \& {Evans}, N.~W. 2024, \mnras, 529, L73

\bibitem[{{de Boer} {et~al.}(2014){de Boer}, {Belokurov}, {Beers}, \& {Lee}}]{deboer14}
{de Boer}, T.~J.~L., {Belokurov}, V., {Beers}, T.~C., \& {Lee}, Y.~S. 2014, \mnras, 443, 658

\bibitem[{{de Boer} {et~al.}(2015){de Boer}, {Belokurov}, \& {Koposov}}]{deboer15}
{de Boer}, T.~J.~L., {Belokurov}, V., \& {Koposov}, S. 2015, \mnras, 451, 3489

\bibitem[{{Dierickx} \& {Loeb}(2017)}]{Dierickx2017}
{Dierickx}, M. I.~P. \& {Loeb}, A. 2017, \apj, 836, 92

\bibitem[{{Fardal} {et~al.}(2019){Fardal}, {van der Marel}, {Law}, {Sohn}, {Sesar}, {Hernitschek}, \& {Rix}}]{Fardal2019}
{Fardal}, M.~A., {van der Marel}, R.~P., {Law}, D.~R., {et~al.} 2019, \mnras, 483, 4724

\bibitem[{{Gaia Collaboration} {et~al.}(2018){Gaia Collaboration}, {Brown}, {Vallenari}, {Prusti}, {de Bruijne}, {Babusiaux}, {Bailer-Jones}, {Biermann}, {Evans}, {Eyer}, {Jansen}, {Jordi}, {Klioner}, {Lammers}, {Lindegren}, {Luri}, {Mignard}, {Panem}, {Pourbaix}, {Randich}, {Sartoretti}, {Siddiqui}, {Soubiran}, {van Leeuwen}, {Walton}, {Arenou}, {Bastian}, {Cropper}, {Drimmel}, {Katz}, {Lattanzi}, {Bakker}, {Cacciari}, {Casta{\~n}eda}, {Chaoul}, {Cheek}, {De Angeli}, {Fabricius}, {Guerra}, {Holl}, {Masana}, {Messineo}, {Mowlavi}, {Nienartowicz}, {Panuzzo}, {Portell}, {Riello}, {Seabroke}, {Tanga}, {Th{\'e}venin}, {Gracia-Abril}, {Comoretto}, {Garcia-Reinaldos}, {Teyssier}, {Altmann}, {Andrae}, {Audard}, {Bellas-Velidis}, {Benson}, {Berthier}, {Blomme}, {Burgess}, {Busso}, {Carry}, {Cellino}, {Clementini}, {Clotet}, {Creevey}, {Davidson}, {De Ridder}, {Delchambre}, {Dell'Oro}, {Ducourant}, {Fern{\'a}ndez-Hern{\'a}ndez}, {Fouesneau}, {Fr{\'e}mat}, {Galluccio}, {Garc{\'\i}a-Torres},
  {Gonz{\'a}lez-N{\'u}{\~n}ez}, {Gonz{\'a}lez-Vidal}, {Gosset}, {Guy}, {Halbwachs}, {Hambly}, {Harrison}, {Hern{\'a}ndez}, {Hestroffer}, {Hodgkin}, {Hutton}, {Jasniewicz}, {Jean-Antoine-Piccolo}, {Jordan}, {Korn}, {Krone-Martins}, {Lanzafame}, {Lebzelter}, {L{\"o}ffler}, {Manteiga}, {Marrese}, {Mart{\'\i}n-Fleitas}, {Moitinho}, {Mora}, {Muinonen}, {Osinde}, {Pancino}, {Pauwels}, {Petit}, {Recio-Blanco}, {Richards}, {Rimoldini}, {Robin}, {Sarro}, {Siopis}, {Smith}, {Sozzetti}, {S{\"u}veges}, {Torra}, {van Reeven}, {Abbas}, {Abreu Aramburu}, {Accart}, {Aerts}, {Altavilla}, {{\'A}lvarez}, {Alvarez}, {Alves}, {Anderson}, {Andrei}, {Anglada Varela}, {Antiche}, {Antoja}, {Arcay}, {Astraatmadja}, {Bach}, {Baker}, {Balaguer-N{\'u}{\~n}ez}, {Balm}, {Barache}, {Barata}, {Barbato}, {Barblan}, {Barklem}, {Barrado}, {Barros}, {Barstow}, {Bartholom{\'e} Mu{\~n}oz}, {Bassilana}, {Becciani}, {Bellazzini}, {Berihuete}, {Bertone}, {Bianchi}, {Bienaym{\'e}}, {Blanco-Cuaresma}, {Boch}, {Boeche}, {Bombrun}, {Borrachero},
  {Bossini}, {Bouquillon}, {Bourda}, {Bragaglia}, {Bramante}, {Breddels}, {Bressan}, {Brouillet}, {Br{\"u}semeister}, {Brugaletta}, {Bucciarelli}, {Burlacu}, {Busonero}, {Butkevich}, {Buzzi}, {Caffau}, {Cancelliere}, {Cannizzaro}, {Cantat-Gaudin}, {Carballo}, {Carlucci}, {Carrasco}, {Casamiquela}, {Castellani}, {Castro-Ginard}, {Charlot}, {Chemin}, {Chiavassa}, {Cocozza}, {Costigan}, {Cowell}, {Crifo}, {Crosta}, {Crowley}, {Cuypers}, {Dafonte}, {Damerdji}, {Dapergolas}, {David}, {David}, {de Laverny}, \& {De Luise}}]{Brown2018}
{Gaia Collaboration}, {Brown}, A.~G.~A., {Vallenari}, A., {et~al.} 2018, \aap, 616, A1

\bibitem[{{Gaia Collaboration} {et~al.}(2021){Gaia Collaboration}, {Brown}, {Vallenari}, {Prusti}, {de Bruijne}, {Babusiaux}, {Biermann}, {Creevey}, {Evans}, {Eyer}, {Hutton}, {Jansen}, {Jordi}, {Klioner}, {Lammers}, {Lindegren}, {Luri}, {Mignard}, {Panem}, {Pourbaix}, {Randich}, {Sartoretti}, {Soubiran}, {Walton}, {Arenou}, {Bailer-Jones}, {Bastian}, {Cropper}, {Drimmel}, {Katz}, {Lattanzi}, {van Leeuwen}, {Bakker}, {Cacciari}, {Casta{\~n}eda}, {De Angeli}, {Ducourant}, {Fabricius}, {Fouesneau}, {Fr{\'e}mat}, {Guerra}, {Guerrier}, {Guiraud}, {Jean-Antoine Piccolo}, {Masana}, {Messineo}, {Mowlavi}, {Nicolas}, {Nienartowicz}, {Pailler}, {Panuzzo}, {Riclet}, {Roux}, {Seabroke}, {Sordo}, {Tanga}, {Th{\'e}venin}, {Gracia-Abril}, {Portell}, {Teyssier}, {Altmann}, {Andrae}, {Bellas-Velidis}, {Benson}, {Berthier}, {Blomme}, {Brugaletta}, {Burgess}, {Busso}, {Carry}, {Cellino}, {Cheek}, {Clementini}, {Damerdji}, {Davidson}, {Delchambre}, {Dell'Oro}, {Fern{\'a}ndez-Hern{\'a}ndez}, {Galluccio}, {Garc{\'\i}a-Lario},
  {Garcia-Reinaldos}, {Gonz{\'a}lez-N{\'u}{\~n}ez}, {Gosset}, {Haigron}, {Halbwachs}, {Hambly}, {Harrison}, {Hatzidimitriou}, {Heiter}, {Hern{\'a}ndez}, {Hestroffer}, {Hodgkin}, {Holl}, {Jan{\ss}en}, {Jevardat de Fombelle}, {Jordan}, {Krone-Martins}, {Lanzafame}, {L{\"o}ffler}, {Lorca}, {Manteiga}, {Marchal}, {Marrese}, {Moitinho}, {Mora}, {Muinonen}, {Osborne}, {Pancino}, {Pauwels}, {Petit}, {Recio-Blanco}, {Richards}, {Riello}, {Rimoldini}, {Robin}, {Roegiers}, {Rybizki}, {Sarro}, {Siopis}, {Smith}, {Sozzetti}, {Ulla}, {Utrilla}, {van Leeuwen}, {van Reeven}, {Abbas}, {Abreu Aramburu}, {Accart}, {Aerts}, {Aguado}, {Ajaj}, {Altavilla}, {{\'A}lvarez}, {{\'A}lvarez Cid-Fuentes}, {Alves}, {Anderson}, {Anglada Varela}, {Antoja}, {Audard}, {Baines}, {Baker}, {Balaguer-N{\'u}{\~n}ez}, {Balbinot}, {Balog}, {Barache}, {Barbato}, {Barros}, {Barstow}, {Bartolom{\'e}}, {Bassilana}, {Bauchet}, {Baudesson-Stella}, {Becciani}, {Bellazzini}, {Bernet}, {Bertone}, {Bianchi}, {Blanco-Cuaresma}, {Boch}, {Bombrun}, {Bossini},
  {Bouquillon}, {Bragaglia}, {Bramante}, {Breedt}, {Bressan}, {Brouillet}, {Bucciarelli}, {Burlacu}, {Busonero}, {Butkevich}, {Buzzi}, {Caffau}, {Cancelliere}, {C{\'a}novas}, {Cantat-Gaudin}, {Carballo}, {Carlucci}, {Carnerero}, {Carrasco}, {Casamiquela}, {Castellani}, {Castro-Ginard}, {Castro Sampol}, {Chaoul}, {Charlot}, {Chemin}, {Chiavassa}, {Cioni}, {Comoretto}, {Cooper}, {Cornez}, {Cowell}, {Crifo}, {Crosta}, {Crowley}, {Dafonte}, {Dapergolas}, {David}, {David}, {de Laverny}, {De Luise}, {De March}, {De Ridder}, {de Souza}, {de Teodoro}, {de Torres}, {del Peloso}, {del Pozo}, {Delbo}, {Delgado}, {Delgado}, {Delisle}, {Di Matteo}, {Diakite}, {Diener}, {Distefano}, {Dolding}, {Eappachen}, {Edvardsson}, {Enke}, {Esquej}, {Fabre}, {Fabrizio}, {Faigler}, {Fedorets}, {Fernique}, {Fienga}, {Figueras}, {Fouron}, {Fragkoudi}, {Fraile}, {Franke}, {Gai}, {Garabato}, {Garcia-Gutierrez}, {Garc{\'\i}a-Torres}, {Garofalo}, {Gavras}, {Gerlach}, {Geyer}, {Giacobbe}, {Gilmore}, {Girona}, {Giuffrida}, {Gomel}, {Gomez},
  {Gonzalez-Santamaria}, {Gonz{\'a}lez-Vidal}, {Granvik}, {Guti{\'e}rrez-S{\'a}nchez}, {Guy}, {Hauser}, {Haywood}, {Helmi}, {Hidalgo}, {Hilger}, {H{\l}adczuk}, {Hobbs}, {Holland}, {Huckle}, {Jasniewicz}, {Jonker}, {Juaristi Campillo}, {Julbe}, {Karbevska}, {Kervella}, {Khanna}, {Kochoska}, {Kontizas}, {Kordopatis}, {Korn}, {Kostrzewa-Rutkowska}, {Kruszy{\'n}ska}, {Lambert}, {Lanza}, {Lasne}, {Le Campion}, {Le Fustec}, {Lebreton}, {Lebzelter}, {Leccia}, {Leclerc}, {Lecoeur-Taibi}, {Liao}, {Licata}, {Lindstr{\o}m}, {Lister}, {Livanou}, {Lobel}, {Madrero Pardo}, {Managau}, {Mann}, {Marchant}, {Marconi}, {Marcos Santos}, {Marinoni}, {Marocco}, {Marshall}, {Martin Polo}, {Mart{\'\i}n-Fleitas}, {Masip}, {Massari}, {Mastrobuono-Battisti}, {Mazeh}, {McMillan}, {Messina}, {Michalik}, {Millar}, {Mints}, {Molina}, {Molinaro}, {Moln{\'a}r}, {Montegriffo}, {Mor}, {Morbidelli}, {Morel}, {Morris}, {Mulone}, {Munoz}, {Muraveva}, {Murphy}, {Musella}, {Noval}, {Ord{\'e}novic}, {Orr{\`u}}, {Osinde}, {Pagani}, {Pagano},
  {Palaversa}, {Palicio}, {Panahi}, {Pawlak}, {Pe{\~n}alosa Esteller}, {Penttil{\"a}}, {Piersimoni}, {Pineau}, {Plachy}, {Plum}, {Poggio}, {Poretti}, {Poujoulet}, {Pr{\v{s}}a}, {Pulone}, {Racero}, {Ragaini}, {Rainer}, {Raiteri}, {Rambaux}, {Ramos}, {Ramos-Lerate}, {Re Fiorentin}, {Regibo}, {Reyl{\'e}}, {Ripepi}, {Riva}, {Rixon}, {Robichon}, {Robin}, {Roelens}, {Rohrbasser}, {Romero-G{\'o}mez}, {Rowell}, {Royer}, {Rybicki}, {Sadowski}, {Sagrist{\`a} Sell{\'e}s}, {Sahlmann}, {Salgado}, {Salguero}, {Samaras}, {Sanchez Gimenez}, {Sanna}, {Santove{\~n}a}, {Sarasso}, {Schultheis}, {Sciacca}, {Segol}, {Segovia}, {S{\'e}gransan}, {Semeux}, {Shahaf}, {Siddiqui}, {Siebert}, {Siltala}, {Slezak}, {Smart}, {Solano}, {Solitro}, {Souami}, {Souchay}, {Spagna}, {Spoto}, {Steele}, {Steidelm{\"u}ller}, {Stephenson}, {S{\"u}veges}, {Szabados}, {Szegedi-Elek}, {Taris}, {Tauran}, {Taylor}, {Teixeira}, {Thuillot}, {Tonello}, {Torra}, {Torra}, {Turon}, {Unger}, {Vaillant}, {van Dillen}, {Vanel}, {Vecchiato}, {Viala}, {Vicente},
  {Voutsinas}, {Weiler}, {Wevers}, {Wyrzykowski}, {Yoldas}, {Yvard}, {Zhao}, {Zorec}, {Zucker}, {Zurbach}, \& {Zwitter}}]{Brown2021}
{Gaia Collaboration}, {Brown}, A.~G.~A., {Vallenari}, A., {et~al.} 2021, \aap, 649, A1

\bibitem[{{Gaia Collaboration} {et~al.}(2023){Gaia Collaboration}, {Vallenari}, {Brown}, {Prusti}, {de Bruijne}, {Arenou}, {Babusiaux}, {Biermann}, {Creevey}, {Ducourant}, {Evans}, {Eyer}, {Guerra}, {Hutton}, {Jordi}, {Klioner}, {Lammers}, {Lindegren}, {Luri}, {Mignard}, {Panem}, {Pourbaix}, {Randich}, {Sartoretti}, {Soubiran}, {Tanga}, {Walton}, {Bailer-Jones}, {Bastian}, {Drimmel}, {Jansen}, {Katz}, {Lattanzi}, {van Leeuwen}, {Bakker}, {Cacciari}, {Casta{\~n}eda}, {De Angeli}, {Fabricius}, {Fouesneau}, {Fr{\'e}mat}, {Galluccio}, {Guerrier}, {Heiter}, {Masana}, {Messineo}, {Mowlavi}, {Nicolas}, {Nienartowicz}, {Pailler}, {Panuzzo}, {Riclet}, {Roux}, {Seabroke}, {Sordo}, {Th{\'e}venin}, {Gracia-Abril}, {Portell}, {Teyssier}, {Altmann}, {Andrae}, {Audard}, {Bellas-Velidis}, {Benson}, {Berthier}, {Blomme}, {Burgess}, {Busonero}, {Busso}, {C{\'a}novas}, {Carry}, {Cellino}, {Cheek}, {Clementini}, {Damerdji}, {Davidson}, {de Teodoro}, {Nu{\~n}ez Campos}, {Delchambre}, {Dell'Oro}, {Esquej},
  {Fern{\'a}ndez-Hern{\'a}ndez}, {Fraile}, {Garabato}, {Garc{\'\i}a-Lario}, {Gosset}, {Haigron}, {Halbwachs}, {Hambly}, {Harrison}, {Hern{\'a}ndez}, {Hestroffer}, {Hodgkin}, {Holl}, {Jan{\ss}en}, {Jevardat de Fombelle}, {Jordan}, {Krone-Martins}, {Lanzafame}, {L{\"o}ffler}, {Marchal}, {Marrese}, {Moitinho}, {Muinonen}, {Osborne}, {Pancino}, {Pauwels}, {Recio-Blanco}, {Reyl{\'e}}, {Riello}, {Rimoldini}, {Roegiers}, {Rybizki}, {Sarro}, {Siopis}, {Smith}, {Sozzetti}, {Utrilla}, {van Leeuwen}, {Abbas}, {{\'A}brah{\'a}m}, {Abreu Aramburu}, {Aerts}, {Aguado}, {Ajaj}, {Aldea-Montero}, {Altavilla}, {{\'A}lvarez}, {Alves}, {Anders}, {Anderson}, {Anglada Varela}, {Antoja}, {Baines}, {Baker}, {Balaguer-N{\'u}{\~n}ez}, {Balbinot}, {Balog}, {Barache}, {Barbato}, {Barros}, {Barstow}, {Bartolom{\'e}}, {Bassilana}, {Bauchet}, {Becciani}, {Bellazzini}, {Berihuete}, {Bernet}, {Bertone}, {Bianchi}, {Binnenfeld}, {Blanco-Cuaresma}, {Blazere}, {Boch}, {Bombrun}, {Bossini}, {Bouquillon}, {Bragaglia}, {Bramante}, {Breedt},
  {Bressan}, {Brouillet}, {Brugaletta}, {Bucciarelli}, {Burlacu}, {Butkevich}, {Buzzi}, {Caffau}, {Cancelliere}, {Cantat-Gaudin}, {Carballo}, {Carlucci}, {Carnerero}, {Carrasco}, {Casamiquela}, {Castellani}, {Castro-Ginard}, {Chaoul}, {Charlot}, {Chemin}, {Chiaramida}, {Chiavassa}, {Chornay}, {Comoretto}, {Contursi}, {Cooper}, {Cornez}, {Cowell}, {Crifo}, {Cropper}, {Crosta}, {Crowley}, {Dafonte}, {Dapergolas}, {David}, {David}, {de Laverny}, {De Luise}, {De March}, {De Ridder}, {de Souza}, {de Torres}, {del Peloso}, {del Pozo}, {Delbo}, {Delgado}, {Delisle}, {Demouchy}, {Dharmawardena}, {Di Matteo}, {Diakite}, {Diener}, {Distefano}, {Dolding}, {Edvardsson}, {Enke}, {Fabre}, {Fabrizio}, {Faigler}, {Fedorets}, {Fernique}, {Fienga}, {Figueras}, {Fournier}, {Fouron}, {Fragkoudi}, {Gai}, {Garcia-Gutierrez}, {Garcia-Reinaldos}, {Garc{\'\i}a-Torres}, {Garofalo}, {Gavel}, {Gavras}, {Gerlach}, {Geyer}, {Giacobbe}, {Gilmore}, {Girona}, {Giuffrida}, {Gomel}, {Gomez}, {Gonz{\'a}lez-N{\'u}{\~n}ez},
  {Gonz{\'a}lez-Santamar{\'\i}a}, {Gonz{\'a}lez-Vidal}, {Granvik}, {Guillout}, {Guiraud}, {Guti{\'e}rrez-S{\'a}nchez}, {Guy}, {Hatzidimitriou}, {Hauser}, {Haywood}, {Helmer}, {Helmi}, {Sarmiento}, {Hidalgo}, {Hilger}, {H{\l}adczuk}, {Hobbs}, {Holland}, {Huckle}, {Jardine}, {Jasniewicz}, {Jean-Antoine Piccolo}, {Jim{\'e}nez-Arranz}, {Jorissen}, {Juaristi Campillo}, {Julbe}, {Karbevska}, {Kervella}, {Khanna}, {Kontizas}, {Kordopatis}, {Korn}, {K{\'o}sp{\'a}l}, {Kostrzewa-Rutkowska}, {Kruszy{\'n}ska}, {Kun}, {Laizeau}, {Lambert}, {Lanza}, {Lasne}, {Le Campion}, {Lebreton}, {Lebzelter}, {Leccia}, {Leclerc}, {Lecoeur-Taibi}, {Liao}, {Licata}, {Lindstr{\o}m}, {Lister}, {Livanou}, {Lobel}, {Lorca}, {Loup}, {Madrero Pardo}, {Magdaleno Romeo}, {Managau}, {Mann}, {Manteiga}, {Marchant}, {Marconi}, {Marcos}, {Marcos Santos}, {Mar{\'\i}n Pina}, {Marinoni}, {Marocco}, {Marshall}, {Martin Polo}, {Mart{\'\i}n-Fleitas}, {Marton}, {Mary}, {Masip}, {Massari}, {Mastrobuono-Battisti}, {Mazeh}, {McMillan}, {Messina}, {Michalik},
  {Millar}, {Mints}, {Molina}, {Molinaro}, {Moln{\'a}r}, {Monari}, {Mongui{\'o}}, {Montegriffo}, {Montero}, {Mor}, {Mora}, {Morbidelli}, {Morel}, {Morris}, {Muraveva}, {Murphy}, {Musella}, {Nagy}, {Noval}, {Oca{\~n}a}, {Ogden}, {Ordenovic}, {Osinde}, {Pagani}, {Pagano}, {Palaversa}, {Palicio}, {Pallas-Quintela}, {Panahi}, {Payne-Wardenaar}, {Pe{\~n}alosa Esteller}, {Penttil{\"a}}, {Pichon}, {Piersimoni}, {Pineau}, {Plachy}, {Plum}, {Poggio}, {Pr{\v{s}}a}, {Pulone}, {Racero}, {Ragaini}, {Rainer}, {Raiteri}, {Rambaux}, {Ramos}, {Ramos-Lerate}, {Re Fiorentin}, {Regibo}, {Richards}, {Rios Diaz}, {Ripepi}, {Riva}, {Rix}, {Rixon}, {Robichon}, {Robin}, {Robin}, {Roelens}, {Rogues}, {Rohrbasser}, {Romero-G{\'o}mez}, {Rowell}, {Royer}, {Ruz Mieres}, {Rybicki}, {Sadowski}, {S{\'a}ez N{\'u}{\~n}ez}, {Sagrist{\`a} Sell{\'e}s}, {Sahlmann}, {Salguero}, {Samaras}, {Sanchez Gimenez}, {Sanna}, {Santove{\~n}a}, {Sarasso}, {Schultheis}, {Sciacca}, {Segol}, {Segovia}, {S{\'e}gransan}, {Semeux}, {Shahaf}, {Siddiqui}, {Siebert},
  {Siltala}, {Silvelo}, {Slezak}, {Slezak}, {Smart}, {Snaith}, {Solano}, {Solitro}, {Souami}, {Souchay}, {Spagna}, {Spina}, {Spoto}, {Steele}, {Steidelm{\"u}ller}, {Stephenson}, {S{\"u}veges}, {Surdej}, {Szabados}, {Szegedi-Elek}, {Taris}, {Taylor}, {Teixeira}, {Tolomei}, {Tonello}, {Torra}, {Torra}, {Torralba Elipe}, {Trabucchi}, {Tsounis}, {Turon}, {Ulla}, {Unger}, {Vaillant}, {van Dillen}, {van Reeven}, {Vanel}, {Vecchiato}, {Viala}, {Vicente}, {Voutsinas}, {Weiler}, {Wevers}, {Wyrzykowski}, {Yoldas}, {Yvard}, {Zhao}, {Zorec}, {Zucker}, \& {Zwitter}}]{Vallenari2023}
{Gaia Collaboration}, {Vallenari}, A., {Brown}, A.~G.~A., {et~al.} 2023, \aap, 674, A1

\bibitem[{{Garofalo} {et~al.}(2022){Garofalo}, {Delgado}, {Sarro}, {Clementini}, {Muraveva}, {Marconi}, \& {Ripepi}}]{Garofalo2022}
{Garofalo}, A., {Delgado}, H.~E., {Sarro}, L.~M., {et~al.} 2022, \mnras, 513, 788

\bibitem[{{Gibbons} {et~al.}(2017){Gibbons}, {Belokurov}, \& {Evans}}]{gibbons17}
{Gibbons}, S.~L.~J., {Belokurov}, V., \& {Evans}, N.~W. 2017, \mnras, 464, 794

\bibitem[{{GRAVITY Collaboration} {et~al.}(2018){GRAVITY Collaboration}, {Abuter}, {Amorim}, {Anugu}, {Baub{\"o}ck}, {Benisty}, {Berger}, {Blind}, {Bonnet}, {Brandner}, {Buron}, {Collin}, {Chapron}, {Cl{\'e}net}, {Coud{\'e} Du Foresto}, {de Zeeuw}, {Deen}, {Delplancke-Str{\"o}bele}, {Dembet}, {Dexter}, {Duvert}, {Eckart}, {Eisenhauer}, {Finger}, {F{\"o}rster Schreiber}, {F{\'e}dou}, {Garcia}, {Garcia Lopez}, {Gao}, {Gendron}, {Genzel}, {Gillessen}, {Gordo}, {Habibi}, {Haubois}, {Haug}, {Hau{\ss}mann}, {Henning}, {Hippler}, {Horrobin}, {Hubert}, {Hubin}, {Jimenez Rosales}, {Jochum}, {Jocou}, {Kaufer}, {Kellner}, {Kendrew}, {Kervella}, {Kok}, {Kulas}, {Lacour}, {Lapeyr{\`e}re}, {Lazareff}, {Le Bouquin}, {L{\'e}na}, {Lippa}, {Lenzen}, {M{\'e}rand}, {M{\"u}ler}, {Neumann}, {Ott}, {Palanca}, {Paumard}, {Pasquini}, {Perraut}, {Perrin}, {Pfuhl}, {Plewa}, {Rabien}, {Ram{\'\i}rez}, {Ramos}, {Rau}, {Rodr{\'\i}guez-Coira}, {Rohloff}, {Rousset}, {Sanchez-Bermudez}, {Scheithauer}, {Sch{\"o}ller}, {Schuler}, {Spyromilio},
  {Straub}, {Straubmeier}, {Sturm}, {Tacconi}, {Tristram}, {Vincent}, {von Fellenberg}, {Wank}, {Waisberg}, {Widmann}, {Wieprecht}, {Wiest}, {Wiezorrek}, {Woillez}, {Yazici}, {Ziegler}, \& {Zins}}]{Gravity2018}
{GRAVITY Collaboration}, {Abuter}, R., {Amorim}, A., {et~al.} 2018, \aap, 615, L15

\bibitem[{{Hayes} {et~al.}(2020){Hayes}, {Majewski}, {Hasselquist}, {Anguiano}, {Shetrone}, {Law}, {Schiavon}, {Cunha}, {Smith}, {Beaton}, {Price-Whelan}, {Allende Prieto}, {Battaglia}, {Bizyaev}, {Brownstein}, {Cohen}, {Frinchaboy}, {Garc{\'\i}a-Hern{\'a}ndez}, {Lacerna}, {Lane}, {M{\'e}sz{\'a}ros}, {Bidin}, {M{\~{u}}noz}, {Nidever}, {Oravetz}, {Oravetz}, {Pan}, {Roman-Lopes}, {Sobeck}, \& {Stringfellow}}]{Hayes2020}
{Hayes}, C.~R., {Majewski}, S.~R., {Hasselquist}, S., {et~al.} 2020, \apj, 889, 63

\bibitem[{{Helmi} \& {White}(2001)}]{Helmi2001}
{Helmi}, A. \& {White}, S. D.~M. 2001, \mnras, 323, 529

\bibitem[{{Hernitschek} {et~al.}(2017){Hernitschek}, {Sesar}, {Rix}, {Belokurov}, {Martinez-Delgado}, {Martin}, {Kaiser}, {Hodapp}, {Chambers}, {Wainscoat}, {Magnier}, {Kudritzki}, {Metcalfe}, \& {Draper}}]{Hernitschek2017}
{Hernitschek}, N., {Sesar}, B., {Rix}, H.-W., {et~al.} 2017, \apj, 850, 96

\bibitem[{{Ibata} {et~al.}(2020){Ibata}, {Bellazzini}, {Thomas}, {Malhan}, {Martin}, {Famaey}, \& {Siebert}}]{Ibata2020}
{Ibata}, R., {Bellazzini}, M., {Thomas}, G., {et~al.} 2020, \apjl, 891, L19

\bibitem[{{Ibata} {et~al.}(1994){Ibata}, {Gilmore}, \& {Irwin}}]{Ibata1994}
{Ibata}, R.~A., {Gilmore}, G., \& {Irwin}, M.~J. 1994, \nat, 370, 194

\bibitem[{{Ibata} \& {Lewis}(1998)}]{Ibata1998}
{Ibata}, R.~A. \& {Lewis}, G.~F. 1998, \apj, 500, 575

\bibitem[{{Jurcsik} \& {Kovacs}(1996)}]{Jurcsik1996}
{Jurcsik}, J. \& {Kovacs}, G. 1996, \aap, 312, 111

\bibitem[{{Kaiser} {et~al.}(2010){Kaiser}, {Burgett}, {Chambers}, {Denneau}, {Heasley}, {Jedicke}, {Magnier}, {Morgan}, {Onaka}, \& {Tonry}}]{Kaiser2010}
{Kaiser}, N., {Burgett}, W., {Chambers}, K., {et~al.} 2010, in Society of Photo-Optical Instrumentation Engineers (SPIE) Conference Series, Vol. 7733, Ground-based and Airborne Telescopes III, ed. L.~M. {Stepp}, R.~{Gilmozzi}, \& H.~J. {Hall}, 77330E

\bibitem[{{Koposov} {et~al.}(2012){Koposov}, {Belokurov}, {Evans}, {Gilmore}, {Gieles}, {Irwin}, {Lewis}, {Niederste-Ostholt}, {Pe{\~n}arrubia}, {Smith}, {Bizyaev}, {Malanushenko}, {Malanushenko}, {Schneider}, \& {Wyse}}]{Koposov2012}
{Koposov}, S.~E., {Belokurov}, V., {Evans}, N.~W., {et~al.} 2012, \apj, 750, 80

\bibitem[{{Law} \& {Majewski}(2010)}]{Law2010}
{Law}, D.~R. \& {Majewski}, S.~R. 2010, \apj, 714, 229

\bibitem[{{Li} {et~al.}(2023){Li}, {Huang}, {Liu}, {Beers}, \& {Zhang}}]{Li2023}
{Li}, X.-Y., {Huang}, Y., {Liu}, G.-C., {Beers}, T.~C., \& {Zhang}, H.-W. 2023, \apj, 944, 88

\bibitem[{{Limberg} {et~al.}(2023){Limberg}, {Queiroz}, {Perottoni}, {Rossi}, {Amarante}, {Santucci}, {Chiappini}, {P{\'e}rez-Villegas}, \& {Lee}}]{limberg23}
{Limberg}, G., {Queiroz}, A. B.~A., {Perottoni}, H.~D., {et~al.} 2023, \apj, 946, 66

\bibitem[{{Liu} {et~al.}(2020){Liu}, {Huang}, {Zhang}, {Xiang}, {Ren}, {Chen}, {Yuan}, {Wang}, {Yang}, {Tian}, {Wang}, \& {Liu}}]{Liu2020}
{Liu}, G.~C., {Huang}, Y., {Zhang}, H.~W., {et~al.} 2020, \apjs, 247, 68

\bibitem[{{Longmore} {et~al.}(1986){Longmore}, {Fernley}, \& {Jameson}}]{Longmore1986}
{Longmore}, A.~J., {Fernley}, J.~A., \& {Jameson}, R.~F. 1986, \mnras, 220, 279

\bibitem[{{Madore} {et~al.}(2013){Madore}, {Hoffman}, {Freedman}, {Kollmeier}, {Monson}, {Persson}, {Rich}, {Scowcroft}, \& {Seibert}}]{Madore2013}
{Madore}, B.~F., {Hoffman}, D., {Freedman}, W.~L., {et~al.} 2013, \apj, 776, 135

\bibitem[{{Majewski} {et~al.}(2003){Majewski}, {Skrutskie}, {Weinberg}, \& {Ostheimer}}]{Majewski2003}
{Majewski}, S.~R., {Skrutskie}, M.~F., {Weinberg}, M.~D., \& {Ostheimer}, J.~C. 2003, \apj, 599, 1082

\bibitem[{{Minelli} {et~al.}(2023){Minelli}, {Bellazzini}, {Mucciarelli}, {Bonifacio}, {Ibata}, {Romano}, {Monaco}, {Caffau}, {Dalessandro}, \& {Pascale}}]{minelli23}
{Minelli}, A., {Bellazzini}, M., {Mucciarelli}, A., {et~al.} 2023, \aap, 669, A54

\bibitem[{{Monaco} {et~al.}(2007){Monaco}, {Bellazzini}, {Bonifacio}, {Buzzoni}, {Ferraro}, {Marconi}, {Sbordone}, \& {Zaggia}}]{monaco07}
{Monaco}, L., {Bellazzini}, M., {Bonifacio}, P., {et~al.} 2007, \aap, 464, 201

\bibitem[{{Morgan} {et~al.}(2007){Morgan}, {Wahl}, \& {Wieckhorst}}]{Morgan2007}
{Morgan}, S.~M., {Wahl}, J.~N., \& {Wieckhorst}, R.~M. 2007, \mnras, 374, 1421

\bibitem[{{Mullen} {et~al.}(2021){Mullen}, {Marengo}, {Mart{\'\i}nez-V{\'a}zquez}, {Neeley}, {Bono}, {Dall'Ora}, {Chaboyer}, {Th{\'e}venin}, {Braga}, {Crestani}, {Fabrizio}, {Fiorentino}, {Gilligan}, {Monelli}, \& {Stetson}}]{Mullen2021}
{Mullen}, J.~P., {Marengo}, M., {Mart{\'\i}nez-V{\'a}zquez}, C.~E., {et~al.} 2021, \apj, 912, 144

\bibitem[{{Muraveva} {et~al.}(2018){Muraveva}, {Delgado}, {Clementini}, {Sarro}, \& {Garofalo}}]{Muraveva2018}
{Muraveva}, T., {Delgado}, H.~E., {Clementini}, G., {Sarro}, L.~M., \& {Garofalo}, A. 2018, \mnras, 481, 1195

\bibitem[{{Muraveva} {et~al.}(2025){Muraveva}, {Giannetti}, {Clementini}, {Garofalo}, \& {Monti}}]{Muraveva2025}
{Muraveva}, T., {Giannetti}, A., {Clementini}, G., {Garofalo}, A., \& {Monti}, L. 2025, \mnras, 536, 2749

\bibitem[{{Muraveva} {et~al.}(2015){Muraveva}, {Palmer}, {Clementini}, {Luri}, {Cioni}, {Moretti}, {Marconi}, {Ripepi}, \& {Rubele}}]{Muraveva2015}
{Muraveva}, T., {Palmer}, M., {Clementini}, G., {et~al.} 2015, \apj, 807, 127

\bibitem[{{Navarrete} {et~al.}(2017){Navarrete}, {Belokurov}, {Koposov}, {Irwin}, {Catelan}, {Duffau}, \& {Drake}}]{Navarrete2017}
{Navarrete}, C., {Belokurov}, V., {Koposov}, S.~E., {et~al.} 2017, \mnras, 467, 1329

\bibitem[{{Neeley} {et~al.}(2017){Neeley}, {Marengo}, {Bono}, {Braga}, {Dall'Ora}, {Magurno}, {Marconi}, {Trueba}, {Tognelli}, {Prada Moroni}, {Beaton}, {Freedman}, {Madore}, {Monson}, {Scowcroft}, {Seibert}, \& {Stetson}}]{Neeley2017}
{Neeley}, J.~R., {Marengo}, M., {Bono}, G., {et~al.} 2017, \apj, 841, 84

\bibitem[{{Neeley} {et~al.}(2019){Neeley}, {Marengo}, {Freedman}, {Madore}, {Beaton}, {Hatt}, {Hoyt}, {Monson}, {Rich}, {Sarajedini}, {Seibert}, \& {Scowcroft}}]{Neeley2019}
{Neeley}, J.~R., {Marengo}, M., {Freedman}, W.~L., {et~al.} 2019, \mnras, 490, 4254

\bibitem[{{Nemec} {et~al.}(2013){Nemec}, {Cohen}, {Ripepi}, {Derekas}, {Moskalik}, {Sesar}, {Chadid}, \& {Bruntt}}]{Nemec2013}
{Nemec}, J.~M., {Cohen}, J.~G., {Ripepi}, V., {et~al.} 2013, \apj, 773, 181

\bibitem[{{Oria} {et~al.}(2022){Oria}, {Ibata}, {Ramos}, {Famaey}, \& {Errani}}]{Oria2022}
{Oria}, P.-A., {Ibata}, R., {Ramos}, P., {Famaey}, B., \& {Errani}, R. 2022, \apjl, 932, L14

\bibitem[{{Pace}(2024)}]{Pace2024}
{Pace}, A.~B. 2024, arXiv e-prints, arXiv:2411.07424

\bibitem[{{Pe{\~n}arrubia} {et~al.}(2011){Pe{\~n}arrubia}, {Zucker}, {Irwin}, {Hyde}, {Lane}, {Lewis}, {Gilmore}, {Evans}, \& {Belokurov}}]{Penarrubia2011}
{Pe{\~n}arrubia}, J., {Zucker}, D.~B., {Irwin}, M.~J., {et~al.} 2011, \apjl, 727, L2

\bibitem[{Price-Whelan(2017)}]{gala}
Price-Whelan, A.~M. 2017, The Journal of Open Source Software, 2

\bibitem[{{Prudil} {et~al.}(2021){Prudil}, {Hanke}, {Lemasle}, {Crestani}, {Braga}, {Fabrizio}, {Koch-Hansen}, {Bono}, {Grebel}, {Matsunaga}, {Marengo}, {da Silva}, {Dall'Ora}, {Mart{\'\i}nez-V{\'a}zquez}, {Altavilla}, {Lala}, {Chaboyer}, {Ferraro}, {Fiorentino}, {Gilligan}, {Nonino}, \& {Th{\'e}venin}}]{Prudil2021}
{Prudil}, Z., {Hanke}, M., {Lemasle}, B., {et~al.} 2021, \aap, 648, A78

\bibitem[{{Prudil} {et~al.}(2024){Prudil}, {Kunder}, {D{\'e}k{\'a}ny}, \& {Koch-Hansen}}]{Prudil2024}
{Prudil}, Z., {Kunder}, A., {D{\'e}k{\'a}ny}, I., \& {Koch-Hansen}, A.~J. 2024, \aap, 684, A176

\bibitem[{{Ramos} {et~al.}(2022){Ramos}, {Antoja}, {Yuan}, {Arentsen}, {Oria}, {Famaey}, {Ibata}, {Mateu}, \& {Carballo-Bello}}]{Ramos2022}
{Ramos}, P., {Antoja}, T., {Yuan}, Z., {et~al.} 2022, \aap, 666, A64

\bibitem[{{Ramos} {et~al.}(2020){Ramos}, {Mateu}, {Antoja}, {Helmi}, {Castro-Ginard}, {Balbinot}, \& {Carrasco}}]{Ramos2020}
{Ramos}, P., {Mateu}, C., {Antoja}, T., {et~al.} 2020, \aap, 638, A104

\bibitem[{{Rimoldini} {et~al.}(2019){Rimoldini}, {Holl}, {Audard}, {Mowlavi}, {Nienartowicz}, {Evans}, {Guy}, {Lecoeur-Ta{\"\i}bi}, {Jevardat de Fombelle}, {Marchal}, {Roelens}, {De Ridder}, {Sarro}, {Regibo}, {Lopez}, {Clementini}, {Ripepi}, {Molinaro}, {Garofalo}, {Moln{\'a}r}, {Plachy}, {Juh{\'a}sz}, {Szabados}, {Lebzelter}, {Teyssier}, \& {Eyer}}]{Rimoldini2019}
{Rimoldini}, L., {Holl}, B., {Audard}, M., {et~al.} 2019, \aap, 625, A97

\bibitem[{{Ripepi} {et~al.}(2019){Ripepi}, {Molinaro}, {Musella}, {Marconi}, {Leccia}, \& {Eyer}}]{Ripepi2019}
{Ripepi}, V., {Molinaro}, R., {Musella}, I., {et~al.} 2019, \aap, 625, A14

\bibitem[{{Sesar} {et~al.}(2017{\natexlab{a}}){Sesar}, {Hernitschek}, {Dierickx}, {Fardal}, \& {Rix}}]{Sesar2017}
{Sesar}, B., {Hernitschek}, N., {Dierickx}, M. I.~P., {Fardal}, M.~A., \& {Rix}, H.-W. 2017{\natexlab{a}}, \apjl, 844, L4

\bibitem[{{Sesar} {et~al.}(2017{\natexlab{b}}){Sesar}, {Hernitschek}, {Mitrovi{\'c}}, {Ivezi{\'c}}, {Rix}, {Cohen}, {Bernard}, {Grebel}, {Martin}, {Schlafly}, {Burgett}, {Draper}, {Flewelling}, {Kaiser}, {Kudritzki}, {Magnier}, {Metcalfe}, {Tonry}, \& {Waters}}]{Sesar2017b}
{Sesar}, B., {Hernitschek}, N., {Mitrovi{\'c}}, S., {et~al.} 2017{\natexlab{b}}, \aj, 153, 204

\bibitem[{{Sollima} {et~al.}(2008){Sollima}, {Cacciari}, {Arkharov}, {Larionov}, {Gorshanov}, {Efimova}, \& {Piersimoni}}]{Sollima2008}
{Sollima}, A., {Cacciari}, C., {Arkharov}, A.~A.~H., {et~al.} 2008, \mnras, 384, 1583

\bibitem[{{Soszy{\'n}ski} {et~al.}(2014){Soszy{\'n}ski}, {Udalski}, {Szyma{\'n}ski}, {Pietrukowicz}, {Mr{\'o}z}, {Skowron}, {Koz{\l}owski}, {Poleski}, {Skowron}, {Pietrzy{\'n}ski}, {Wyrzykowski}, {Ulaczyk}, \& {Kubiak}}]{OGLE1}
{Soszy{\'n}ski}, I., {Udalski}, A., {Szyma{\'n}ski}, M.~K., {et~al.} 2014, \actaa, 64, 177

\bibitem[{{Soszy{\'n}ski} {et~al.}(2019){Soszy{\'n}ski}, {Udalski}, {Wrona}, {Szyma{\'n}ski}, {Pietrukowicz}, {Skowron}, {Skowron}, {Poleski}, {Koz{\l}owski}, {Mr{\'o}z}, {Ulaczyk}, {Rybicki}, {Iwanek}, \& {Gromadzki}}]{OGLE2}
{Soszy{\'n}ski}, I., {Udalski}, A., {Wrona}, M., {et~al.} 2019, \actaa, 69, 321

\bibitem[{{Sun} {et~al.}(2025){Sun}, {Wang}, {Zhang}, {Xue}, {Huang}, {Zhang}, {Rix}, {Li}, {Liu}, {Zhang}, {Yang}, \& {Zhang}}]{Sun2025}
{Sun}, S., {Wang}, F., {Zhang}, H., {et~al.} 2025, \apj, 979, 213

\bibitem[{{Thomas} {et~al.}(2017){Thomas}, {Famaey}, {Ibata}, {L{\"u}ghausen}, \& {Kroupa}}]{Thomas2017}
{Thomas}, G.~F., {Famaey}, B., {Ibata}, R., {L{\"u}ghausen}, F., \& {Kroupa}, P. 2017, \aap, 603, A65

\bibitem[{{Vasiliev} {et~al.}(2021){Vasiliev}, {Belokurov}, \& {Erkal}}]{Vasiliev2021}
{Vasiliev}, E., {Belokurov}, V., \& {Erkal}, D. 2021, \mnras, 501, 2279

\bibitem[{{Vera-Ciro} \& {Helmi}(2013)}]{Vera-Ciro2013}
{Vera-Ciro}, C. \& {Helmi}, A. 2013, \apjl, 773, L4

\bibitem[{{Wang} {et~al.}(2022){Wang}, {Zhang}, {Xue}, {Huang}, {Liu}, {Zhang}, \& {Yang}}]{Wang2022}
{Wang}, F., {Zhang}, H.~W., {Xue}, X.~X., {et~al.} 2022, \mnras, 513, 1958

\bibitem[{{Yang} {et~al.}(2019){Yang}, {Xue}, {Li}, {Liu}, {Zhang}, {Rix}, {Zhang}, {Zhao}, {Tian}, {Zhong}, {Xing}, {Wu}, {Li}, {Carlin}, \& {Chang}}]{Yang2019}
{Yang}, C., {Xue}, X.-X., {Li}, J., {et~al.} 2019, \apj, 886, 154

\bibitem[{{Yanny} {et~al.}(2009){Yanny}, {Rockosi}, {Newberg}, {Knapp}, {Adelman-McCarthy}, {Alcorn}, {Allam}, {Allende Prieto}, {An}, {Anderson}, {Anderson}, {Bailer-Jones}, {Bastian}, {Beers}, {Bell}, {Belokurov}, {Bizyaev}, {Blythe}, {Bochanski}, {Boroski}, {Brinchmann}, {Brinkmann}, {Brewington}, {Carey}, {Cudworth}, {Evans}, {Evans}, {Gates}, {G{\"a}nsicke}, {Gillespie}, {Gilmore}, {Nebot Gomez-Moran}, {Grebel}, {Greenwell}, {Gunn}, {Jordan}, {Jordan}, {Harding}, {Harris}, {Hendry}, {Holder}, {Ivans}, {Ivezi{\v{c}}}, {Jester}, {Johnson}, {Kent}, {Kleinman}, {Kniazev}, {Krzesinski}, {Kron}, {Kuropatkin}, {Lebedeva}, {Lee}, {French Leger}, {L{\'e}pine}, {Levine}, {Lin}, {Long}, {Loomis}, {Lupton}, {Malanushenko}, {Malanushenko}, {Margon}, {Martinez-Delgado}, {McGehee}, {Monet}, {Morrison}, {Munn}, {Neilsen}, {Nitta}, {Norris}, {Oravetz}, {Owen}, {Padmanabhan}, {Pan}, {Peterson}, {Pier}, {Platson}, {Re Fiorentin}, {Richards}, {Rix}, {Schlegel}, {Schneider}, {Schreiber}, {Schwope}, {Sibley}, {Simmons},
  {Snedden}, {Allyn Smith}, {Stark}, {Stauffer}, {Steinmetz}, {Stoughton}, {SubbaRao}, {Szalay}, {Szkody}, {Thakar}, {Sivarani}, {Tucker}, {Uomoto}, {Vanden Berk}, {Vidrih}, {Wadadekar}, {Watters}, {Wilhelm}, {Wyse}, {Yarger}, \& {Zucker}}]{Yanny2009}
{Yanny}, B., {Rockosi}, C., {Newberg}, H.~J., {et~al.} 2009, \aj, 137, 4377

\bibitem[{{York} {et~al.}(2000){York}, {Adelman}, {Anderson}, {Anderson}, {Annis}, {Bahcall}, {Bakken}, {Barkhouser}, {Bastian}, {Berman}, {Boroski}, {Bracker}, {Briegel}, {Briggs}, {Brinkmann}, {Brunner}, {Burles}, {Carey}, {Carr}, {Castander}, {Chen}, {Colestock}, {Connolly}, {Crocker}, {Csabai}, {Czarapata}, {Davis}, {Doi}, {Dombeck}, {Eisenstein}, {Ellman}, {Elms}, {Evans}, {Fan}, {Federwitz}, {Fiscelli}, {Friedman}, {Frieman}, {Fukugita}, {Gillespie}, {Gunn}, {Gurbani}, {de Haas}, {Haldeman}, {Harris}, {Hayes}, {Heckman}, {Hennessy}, {Hindsley}, {Holm}, {Holmgren}, {Huang}, {Hull}, {Husby}, {Ichikawa}, {Ichikawa}, {Ivezi{\'c}}, {Kent}, {Kim}, {Kinney}, {Klaene}, {Kleinman}, {Kleinman}, {Knapp}, {Korienek}, {Kron}, {Kunszt}, {Lamb}, {Lee}, {Leger}, {Limmongkol}, {Lindenmeyer}, {Long}, {Loomis}, {Loveday}, {Lucinio}, {Lupton}, {MacKinnon}, {Mannery}, {Mantsch}, {Margon}, {McGehee}, {McKay}, {Meiksin}, {Merelli}, {Monet}, {Munn}, {Narayanan}, {Nash}, {Neilsen}, {Neswold}, {Newberg}, {Nichol}, {Nicinski},
  {Nonino}, {Okada}, {Okamura}, {Ostriker}, {Owen}, {Pauls}, {Peoples}, {Peterson}, {Petravick}, {Pier}, {Pope}, {Pordes}, {Prosapio}, {Rechenmacher}, {Quinn}, {Richards}, {Richmond}, {Rivetta}, {Rockosi}, {Ruthmansdorfer}, {Sandford}, {Schlegel}, {Schneider}, {Sekiguchi}, {Sergey}, {Shimasaku}, {Siegmund}, {Smee}, {Smith}, {Snedden}, {Stone}, {Stoughton}, {Strauss}, {Stubbs}, {SubbaRao}, {Szalay}, {Szapudi}, {Szokoly}, {Thakar}, {Tremonti}, {Tucker}, {Uomoto}, {Vanden Berk}, {Vogeley}, {Waddell}, {Wang}, {Watanabe}, {Weinberg}, {Yanny}, {Yasuda}, \& {SDSS Collaboration}}]{York2000}
{York}, D.~G., {Adelman}, J., {Anderson}, Jr., J.~E., {et~al.} 2000, \aj, 120, 1579

\bibitem[{{Zinn} \& {West}(1984)}]{ZW1984}
{Zinn}, R. \& {West}, M.~J. 1984, \apjs, 55, 45

\end{thebibliography}

\begin{appendix}

\section{Datasets}

\begin{table*}
\centering

\caption{Parameters of the 3865 RRLs in the RRLS-SGR-R22 sample.}
\label{tab:rrls_sgr_r22} 
\begin{tabular}{ccccccccc} 
\hline 
source\_id & RA & DEC & \texttt{pmra} & \texttt{pmdec} & [Fe/H] &  $\sigma_{\rm [Fe/H]}$ & $D$ & $\sigma_{D}$ \\
\hline

4047184898945785600 & 279.7316 & -30.1663 & -3.0127 & -1.4424 & -1.41 & 0.59 & 25.07 & 2.37 \\
4047129442313532160 & 279.7830 & -30.6957 & -2.7669 & -1.5504 & -1.61 & 0.42 & 25.06 & 2.24 \\
4047077176843468800 & 279.8683 & -30.8780 & -2.7674 & -1.1131 & -1.49 & 0.48 & 26.50 & 2.54 \\
4047105905883418624 & 279.8829 & -30.6763 & -2.1786 & -1.0593 & -1.61 & 0.50 & 28.77 & 2.99 \\
4047159949451686016 & 279.8989 & -30.4508 & -3.0120 & -1.3126 & -1.46 & 0.54 & 25.42 & 2.39 \\
4047167272398352768 & 279.9083 & -30.3439 & -2.4787 & -1.1172 & -1.60 & 0.43 & 26.69 & 2.17 \\
4047280517778027776 & 279.9369 & -30.0449 & -2.3238 & -1.0802 & -2.01 & 0.49 & 29.66 & 2.31 \\
4047471661003726336 & 279.9499 & -29.8957 & -2.2263 & -1.1029 & -1.82 & 0.43 & 27.26 & 2.39 \\
4047170158616561536 & 279.9889 & -30.2228 & -2.8408 & -1.1482 & -1.73 & 0.47 & 22.56 & 2.27 \\
\hline

\end{tabular} 
\tablefoot{Column (1) {\it Gaia} DR3 source\_id; (2) and (3) Coordinates; (4) and (5) Proper motions; (6) and (7) Photometric metallicities and their corresponding uncertainties; (8) and (9) Heliocentric distances and their corresponding uncertainties, in kpc.
Columns (1)–(5) are taken from the {\it Gaia} DR3 \texttt{gaia\_source} table \citep{Vallenari2023}.
Columns (6) and (7) are from \citet{Muraveva2025}.
Columns (8) and (9) are calculated in this study using the $PWZ$ relation from \citet{Garofalo2022}.
The table is published in its entirety as Supporting Information with the electronic version of the article. A portion is shown here for guidance regarding its form and content.}

\end{table*}

\begin{table*}
\centering

\caption{Parameters of the 2377 RRLs in the RRLS-SGR-MODEL sample.}\label{tab:rrls_sgr_model} 
\begin{tabular}{ccccccccc} 
\hline 
source\_id & RA & DEC & \texttt{pmra} & \texttt{pmdec} & [Fe/H] &  $\sigma_{\rm [Fe/H]}$ & $D$ & $\sigma_{D}$ \\
\hline

4126391959941481344 & 254.6440 &-21.1857 &-1.9582 &-0.4029 &-0.75 &0.69 &31.09 &3.01 \\
4049275993982198784 & 273.0300 &-30.3651 &-2.7870 &-0.8349 &-1.99 &0.52 &26.89 &2.56\\
4109887504462731392 & 260.5566 &-26.1711 &-2.7374 &-1.2626 &-1.74 &0.44 &30.36 &3.36\\
4051354380219168256 & 274.5878 &-28.6183 &-2.6982 &-1.4869 &-1.37 &0.45 &26.23 &2.34\\
4049773458566813312 & 274.7462 &-29.2066 &-3.2204 &-1.5801 &-1.38 &0.59 &26.21 &2.66\\
4049479888952738560 & 275.2280 &-30.2035 &-2.9587 &-1.5748 &-1.74 &0.44 &25.75 &3.78\\
4046444068672593664 & 275.3101 &-30.4965 &-2.6414 &-0.5900 &-1.63 &0.43 &27.54 &2.66\\
4052698979836276864 & 275.3452 &-26.9998 &-2.6180 &-1.3422 &-1.73 &0.58 &26.39 &3.29\\
4051538002998693376 & 275.3472 &-27.6480 &-2.9746 &-1.6238 &-1.31 &0.42 &24.77 &3.33\\

\hline

\end{tabular} 
\tablefoot{The same as in Table~\ref{tab:rrls_sgr_r22}.}

\end{table*} 

\begin{table*}
\centering

\caption{Parameters of the 5296 RRLs in the RRLS-SGR-SELECTION sample.}
\label{tab:rrls_sgr_selection} 
\begin{tabular}{ccccccccc} 
\hline 
source\_id & RA & DEC & \texttt{pmra} & \texttt{pmdec} & [Fe/H] &  $\sigma_{\rm [Fe/H]}$ & $D$ & $\sigma_{D}$ \\
\hline
4126391959941481344 & 254.6440 &-21.1857 &-1.9582 &-0.4029 &-0.75 &0.69 &31.09 &3.01 \\
3278874447192970880 & 68.2296  &  1.4347 &-0.0036 &-1.5532 &-0.83 &0.39 &22.94 &1.83 \\
4049213115752945024 & 272.9878 &-30.7212 &-2.6923 &-1.0830 &-1.72 &0.53 &21.00 &2.70 \\
4049275993982198784 & 273.0300 &-30.3651 &-2.7870 &-0.8349 &-1.99 &0.52 &26.89 &2.56\\
4049745386708273152 & 273.9588 &-29.5741 &-1.6353 &-0.8313 &-2.01 &0.92 &24.13 &3.45\\
4052147746648211712 & 274.2707 &-28.0143 &-2.3168 &-1.7097 &-1.96 &0.45 &25.71 &2.95\\
4051394993395236096 & 274.8139 &-28.3340 &-2.6007 &-1.4730 &-0.93 &0.66 &21.96 &2.97\\
4040907718345693824 & 265.9934 &-36.5574 &-2.3420 &-0.3696 &-1.43 &0.52 &16.96 &1.68\\
4119077703729739520 & 267.1176 &-19.8543 &-2.4453 &-0.6449 &-1.92 &0.53 &16.02 &1.88\\

\hline

\end{tabular} 
\tablefoot{The same as in Table~\ref{tab:rrls_sgr_r22}.}

\end{table*}

\section{Plots}

\begin{figure*}
\includegraphics[width=17cm]{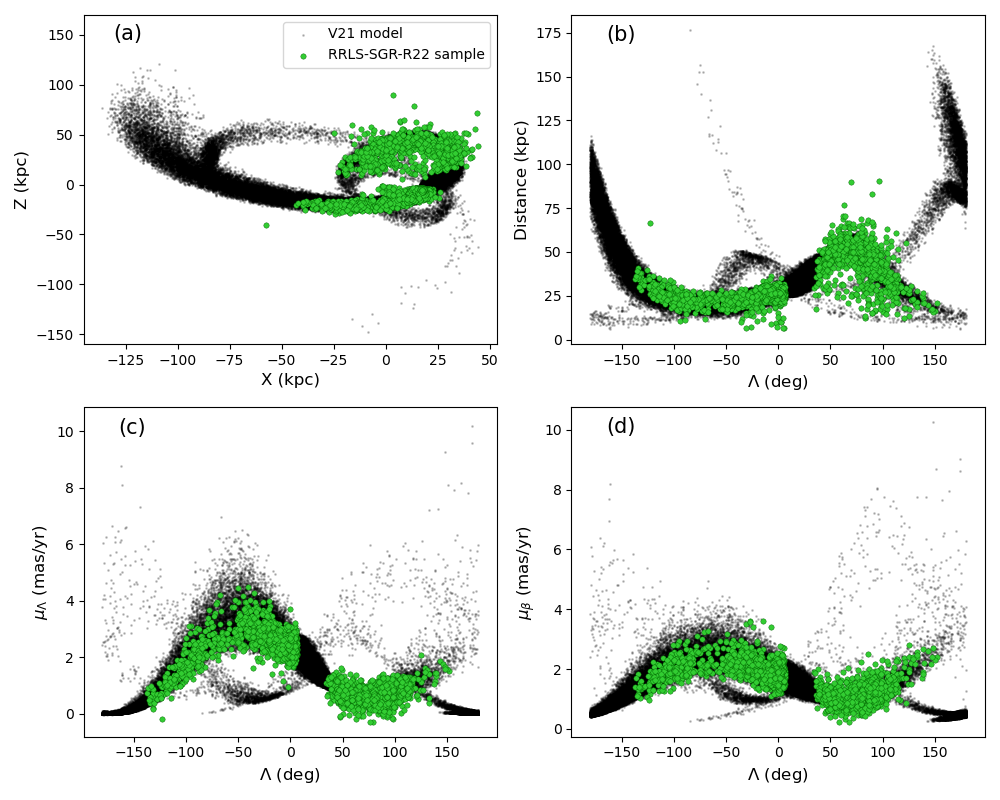}
\caption{The distributions of RRLs from the RRLS-SGR-R22 sample (green points) and simulated stars from the \citet{Vasiliev2021} model (black dots) are shown in the following planes: (a) Cartesian $Z$ versus $X$, (b) distance versus the Sgr $\Lambda$ coordinate, (c) proper motion in the Sgr system $\mu_\Lambda$ versus $\Lambda$, and (d) $\mu_B$ versus $\Lambda$ coordinate. The Sgr coordinates and proper motions are in the system introduced by \citet{Vasiliev2021}.}\label{fig:4panel_ramos}
\end{figure*}

\begin{figure*}
\includegraphics[width=17cm]{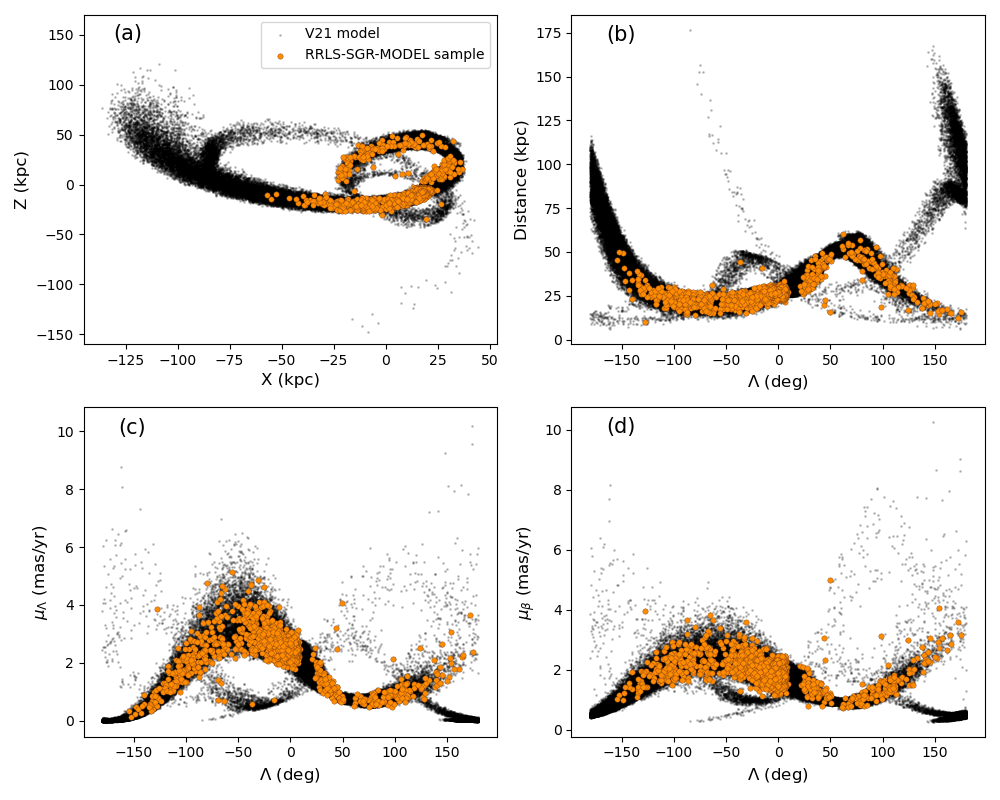}
\caption{The distributions of RRLs from the RRLS-SGR-MODEL sample (orange points) and simulated stars from the \citet{Vasiliev2021} model (black dots) are shown in the following planes: (a) Cartesian $Z$ versus $X$, (b) distance versus the Sgr $\Lambda$ coordinate, (c) proper motion in the Sgr system $\mu_\Lambda$ versus $\Lambda$, and (d) $\mu_B$ versus $\Lambda$ coordinate. The Sgr coordinates and proper motions are in the system introduced by \citet{Vasiliev2021}.}\label{fig:4panel_model}
\end{figure*}

In this section, we present the distributions of RRLs from the RRLS-SGR-R22 (Fig.~\ref{fig:4panel_ramos}) and RRLS-SGR-MODEL (Fig.~\ref{fig:4panel_model}) samples, overlaid on the present-day snapshot from the \citet{Vasiliev2021} model. Panel (a) shows the distributions in the Cartesian $Z$ versus $X$ plane; panel (b) presents the distribution of distances versus the Sgr $\Lambda$ coordinate for the same samples; while panels (c) and (d) display the {\it Gaia} DR3 proper motions, transformed into Sgr coordinates, $\mu_\Lambda$ and $\mu_B$, as functions of $\Lambda$.

\end{appendix}

\end{document}